\expandafter\ifx\csname phyzzx\endcsname\relax
 \message{It is better to use PHYZZX format than to
          \string\input\space PHYZZX}\else
 \wlog{PHYZZX macros are already loaded and are not
          \string\input\space again}%
 \endinput \fi
\catcode`\@=11 
\let\rel@x=\relax
\let\n@expand=\relax
\def\pr@tect{\let\n@expand=\noexpand}
\let\protect=\pr@tect
\let\gl@bal=\global
%
%
%
\newfam\cpfam
\newdimen\b@gheight             \b@gheight=12pt
\newcount\f@ntkey               \f@ntkey=0
\def\f@m{\afterassignment\samef@nt\f@ntkey=}
\def\samef@nt{\fam=\f@ntkey \the\textfont\f@ntkey\rel@x}
\def\setstr@t{\setbox\strutbox=\hbox{\vrule height 0.85\b@gheight
                                depth 0.35\b@gheight width\z@ }}
%
%
%
%
\font\seventeenrm =cmr17
\font\fourteenrm  = cmr10 scaled\magstep2  
\font\twelverm    =cmr12
\font\eightrm     =cmr8
\font\sixrm       =cmr6
%
\font\fourteenbf  =cmbx10 scaled\magstep2  
\font\twelvebf    =cmbx12
\font\eightbf     =cmbx8
\font\sixbf       =cmbx6
\font\seventeeni  =cmmi10 scaled\magstep3    \skewchar\seventeeni='177
\font\fourteeni   =cmmi10 scaled\magstep2     \skewchar\fourteeni='177
\font\twelvei     =cmmi12                      \skewchar\twelvei='177
\font\eighti      =cmmi8                         \skewchar\eighti='177
\font\sixi        =cmmi6                           \skewchar\sixi='177
\font\seventeensy =cmsy10 scaled\magstep3    \skewchar\seventeensy='60
\font\fourteensy  =cmsy10 scaled\magstep2     \skewchar\fourteensy='60
\font\twelvesy    =cmsy10 scaled\magstep1       \skewchar\twelvesy='60
\font\eightsy     =cmsy8                         \skewchar\eightsy='60
\font\sixsy       =cmsy6                           \skewchar\sixsy='60
%
\font\fourteenex  =cmex10  scaled\magstep1  
\font\twelveex  =cmex10 scaled\magstep1
%
\font\fourteensl  =cmsl10 scaled\magstep2  
\font\twelvesl    =cmsl12
\font\eightsl     =cmsl8
%
\font\fourteenit  =cmti10 scaled\magstep2  
\font\twelveit    =cmti12
\font\eightit     =cmti8
\font\sevenit     =cmti7
\font\sixit       =cmti7    
\font\fourteentt  =cmtt10 scaled\magstep2   
\font\twelvett    =cmtt12
\font\eighttt     =cmtt8
\font\fourteencp  =cmcsc10 scaled\magstep2
\font\twelvecp    =cmcsc10 scaled\magstep1
\font\tencp       =cmcsc10
%
%
\def\fourteenf@nts{\relax
    \textfont0=\fourteenrm
                \scriptfont0=\tenrm
                              \scriptscriptfont0=\sevenrm
    \textfont1=\fourteeni
                \scriptfont1=\teni
                              \scriptscriptfont1=\seveni
    \textfont2=\fourteensy
                \scriptfont2=\tensy
                              \scriptscriptfont2=\sevensy
    \textfont3=\fourteenex
                \scriptfont3=\twelveex
                              \scriptscriptfont3=\tenex
    \textfont\itfam=\fourteenit
                     \scriptfont\itfam=\tenit
                                 \scriptscriptfont\itfam=\sevenit
    \textfont\slfam=\fourteensl
                     \scriptfont\slfam=\tensl
                                 \scriptscriptfont\slfam=\sevenrm
    \textfont\bffam=\fourteenbf
                     \scriptfont\bffam=\tenbf
                                 \scriptscriptfont\bffam=\sevenbf
   \textfont\ttfam=\fourteentt
                    \scriptfont\ttfam=\tentt
    \textfont\cpfam=\fourteencp
                     \scriptfont\cpfam\tencp}
\def\twelvef@nts{\relax
    \textfont0=\twelverm
                \scriptfont0=\eightrm
                              \scriptscriptfont0=\sixrm
    \textfont1=\twelvei
                \scriptfont1=\eighti
                              \scriptscriptfont1=\sixi
    \textfont2=\twelvesy
                \scriptfont2=\eightsy
                              \scriptscriptfont2=\sixsy
    \textfont3=\twelveex
                \scriptfont3=\tenex
                              \scriptscriptfont3=\tenex
   \textfont\itfam=\twelveit
                    \scriptfont\itfam=\eightit
                                \scriptscriptfont\itfam=\sixit
   \textfont\slfam=\twelvesl
                    \scriptfont\slfam=\eightsl
    \textfont\bffam=\twelvebf
                     \scriptfont\bffam=\eightbf
                                 \scriptscriptfont\bffam=\sixbf
   \textfont\ttfam=\twelvett
             \scriptfont\ttfam=\eighttt
   \textfont\cpfam=\twelvecp }
\def\tenf@nts{\relax
    \textfont0=\tenrm
                \scriptfont0=\sevenrm
                              \scriptscriptfont0=\fiverm
    \textfont1=\teni
                \scriptfont1=\seveni
                              \scriptscriptfont1=\fivei
   \textfont2=\tensy
               \scriptfont2=\sevensy
                             \scriptscriptfont2=\fivesy
    \textfont3=\tenex
                \scriptfont3=\tenex
                              \scriptscriptfont3=\tenex
    \textfont\itfam=\tenit
              \scriptfont\itfam=\sevenit
    \textfont\slfam=\tensl
                     \scriptfont\slfam=\sevenrm 
    \textfont\bffam=\tenbf
                     \scriptfont\bffam=\sevenbf
                                 \scriptscriptfont\bffam=\fivebf
    \textfont\ttfam=\tentt
    \textfont\cpfam=\tencp }
%
%
\def\smallf@nts{\relax
    \textfont0=\sixrm
                \scriptfont0=\sixrm
                              \scriptscriptfont0=\fiverm
    \textfont1=\sixi
                \scriptfont1=\sixi
                              \scriptscriptfont1=\fivei
   \textfont2=\sixsy
               \scriptfont2=\sixsy
                             \scriptscriptfont2=\fivesy
    \textfont3=\tenex
                \scriptfont3=\tenex
                              \scriptscriptfont3=\tenex
    \textfont\itfam=\sixit
              \scriptfont\itfam=\sixit
    \textfont\slfam=\sixrm
                     \scriptfont\slfam=\sixrm 
    \textfont\bffam=\sixbf
                     \scriptfont\bffam=\sixbf
                                 \scriptscriptfont\bffam=\fivebf
    }
%

\def\rm{\n@expand\f@m0 }
\def\mit{\n@expand\f@m1 }         
\def\cal{\n@expand\f@m2 }
\def\it{\n@expand\f@m\itfam}
\def\sl{\n@expand\f@m\slfam}
\def\bf{\n@expand\f@m\bffam}
\def\tt{\n@expand\f@m\ttfam}
\def\caps{\n@expand\f@m\cpfam}    
\def\em@{\rel@x\ifnum\f@ntkey=0 \it \else
        \ifnum\f@ntkey=\bffam \it \else \rm \fi \fi }
\def\em{\n@expand\em@}
\def\fourteenpoint{\fourteenf@nts \samef@nt \b@gheight=14pt \setstr@t }
\def\twelvepoint{\twelvef@nts \samef@nt \b@gheight=12pt \setstr@t }
\def\tenpoint{\tenf@nts \samef@nt \b@gheight=10pt \setstr@t }
\normalbaselineskip = 20pt plus 0.2pt minus 0.1pt
\normallineskip = 1.5pt plus 0.1pt minus 0.1pt
\normallineskiplimit = 1.5pt
\newskip\normaldisplayskip
\normaldisplayskip = 20pt plus 5pt minus 10pt
\newskip\normaldispshortskip
\normaldispshortskip = 6pt plus 5pt
\newskip\normalparskip
\normalparskip = 6pt plus 2pt minus 1pt
\newskip\skipregister
\skipregister = 5pt plus 2pt minus 1.5pt
\newif\ifsingl@
\newif\ifdoubl@
\newif\iftwelv@  \twelv@true
\def\singlespace{\singl@true\doubl@false\spaces@t}
\def\doublespace{\singl@false\doubl@true\spaces@t}
\def\normalspace{\singl@false\doubl@false\spaces@t}
\def\Tenpoint{\tenpoint\twelv@false\spaces@t}
\def\Twelvepoint{\twelvepoint\twelv@true\spaces@t}
\def\spaces@t{\rel@x
      \iftwelv@ \ifsingl@\subspaces@t3:4;\else\subspaces@t1:1;\fi
       \else \ifsingl@\subspaces@t3:5;\else\subspaces@t4:5;\fi \fi
      \ifdoubl@ \multiply\baselineskip by 5
         \divide\baselineskip by 4 \fi }
\def\subspaces@t#1:#2;{
      \baselineskip = \normalbaselineskip
      \multiply\baselineskip by #1 \divide\baselineskip by #2
      \lineskip = \normallineskip
      \multiply\lineskip by #1 \divide\lineskip by #2
      \lineskiplimit = \normallineskiplimit
      \multiply\lineskiplimit by #1 \divide\lineskiplimit by #2
      \parskip = \normalparskip
      \multiply\parskip by #1 \divide\parskip by #2
      \abovedisplayskip = \normaldisplayskip
      \multiply\abovedisplayskip by #1 \divide\abovedisplayskip by #2
      \belowdisplayskip = \abovedisplayskip
      \abovedisplayshortskip = \normaldispshortskip
      \multiply\abovedisplayshortskip by #1
        \divide\abovedisplayshortskip by #2
      \belowdisplayshortskip = \abovedisplayshortskip
      \advance\belowdisplayshortskip by \belowdisplayskip
      \divide\belowdisplayshortskip by 2
      \smallskipamount = \skipregister
      \multiply\smallskipamount by #1 \divide\smallskipamount by #2
      \medskipamount = \smallskipamount \multiply\medskipamount by 2
      \bigskipamount = \smallskipamount \multiply\bigskipamount by 4 }
\def\normalbaselines{ \baselineskip=\normalbaselineskip
   \lineskip=\normallineskip \lineskiplimit=\normallineskip
   \iftwelv@\else \multiply\baselineskip by 4 \divide\baselineskip by 5
     \multiply\lineskiplimit by 4 \divide\lineskiplimit by 5
     \multiply\lineskip by 4 \divide\lineskip by 5 \fi }
\Twelvepoint  
\interlinepenalty=50
\interfootnotelinepenalty=5000
\predisplaypenalty=9000
\postdisplaypenalty=500
\hfuzz=1pt
\vfuzz=0.2pt
\newdimen\HOFFSET  \HOFFSET=0pt
\newdimen\VOFFSET  \VOFFSET=0pt
\newdimen\HSWING   \HSWING=0pt
\dimen\footins=8in
%
%
%
\newskip\pagebottomfiller
\pagebottomfiller=\z@ plus \z@ minus \z@
\def\pagecontents{
   \ifvoid\topins\else\unvbox\topins\vskip\skip\topins\fi
   \dimen@ = \dp255 \unvbox255
   \vskip\pagebottomfiller
   \ifvoid\footins\else\vskip\skip\footins\footrule\unvbox\footins\fi
   \ifr@ggedbottom \kern-\dimen@ \vfil \fi }
\def\makeheadline{\vbox to 0pt{ \skip@=\topskip
      \advance\skip@ by -12pt \advance\skip@ by -2\normalbaselineskip
      \vskip\skip@ \line{\vbox to 12pt{}\the\headline} \vss
      }\nointerlineskip}
\def\makefootline{\baselineskip = 1.5\normalbaselineskip
                 \line{\the\footline}}
\newif\iffrontpage
\newif\ifp@genum
\def\nopagenumbers{\p@genumfalse}
\def\pagenumbers{\p@genumtrue}
\pagenumbers
\newtoks\paperheadline
\newtoks\paperfootline
\newtoks\letterheadline
\newtoks\letterfootline
\newtoks\letterinfo
\newtoks\date
\paperheadline={\hfil}
\paperfootline={\hss\iffrontpage\else\ifp@genum\tenrm\folio\hss\fi\fi}
\letterheadline{\iffrontpage \hfil \else
    \rm \ifp@genum page~~\folio\fi \hfil\the\date \fi}
\letterfootline={\iffrontpage\the\letterinfo\else\hfil\fi}
\letterinfo={\hfil}
\def\monthname{\rel@x\ifcase\month 0/\or January\or February\or
   March\or April\or May\or June\or July\or August\or September\or
   October\or November\or December\else\number\month/\fi}
\def\today{\monthname~\number\day, \number\year}
\date={\today}
\headline=\paperheadline 
\footline=\paperfootline 
\countdef\pageno=1      \countdef\pagen@=0
\countdef\pagenumber=1  \pagenumber=1
\def\advancepageno{\gl@bal\advance\pagen@ by 1
   \ifnum\pagenumber<0 \gl@bal\advance\pagenumber by -1
    \else\gl@bal\advance\pagenumber by 1 \fi
    \gl@bal\frontpagefalse  \swing@ }
\def\folio{\ifnum\pagenumber<0 \romannumeral-\pagenumber
           \else \number\pagenumber \fi }
\def\swing@{\ifodd\pagenumber \gl@bal\advance\hoffset by -\HSWING
             \else \gl@bal\advance\hoffset by \HSWING \fi }
\def\footrule{\dimen@=\prevdepth\nointerlineskip
   \vbox to 0pt{\vskip -0.25\baselineskip \hrule width 0.35\hsize \vss}
   \prevdepth=\dimen@ }
\let\footnotespecial=\rel@x
\newdimen\footindent
\footindent=24pt
\def\Textindent#1{\noindent\llap{#1\enspace}\ignorespaces}
\def\Vfootnote#1{\insert\footins\bgroup
   \interlinepenalty=\interfootnotelinepenalty \floatingpenalty=20000
   \singl@true\doubl@false\Tenpoint
   \splittopskip=\ht\strutbox \boxmaxdepth=\dp\strutbox
   \leftskip=\footindent \rightskip=\z@skip
   \parindent=0.5\footindent \parfillskip=0pt plus 1fil
   \spaceskip=\z@skip \xspaceskip=\z@skip \footnotespecial
   \Textindent{#1}\footstrut\futurelet\next\fo@t}

\def\vfootnote#1{\Vfootnote{${#1}$}}
\def\footnote#1{\attach{#1}\vfootnote{#1}}

\def\foot{\attach\footsymbolgen\vfootnote{\footsymbol}}
\let\footsymbol=\star
\newcount\lastf@@t           \lastf@@t=-1
\newcount\footsymbolcount    \footsymbolcount=0
\newif\ifPhysRev
\def\footsymbolgen{\bumpfootsymbolcount \generatefootsymbol \footsymbol }
\def\bumpfootsymbolcount{\rel@x
   \iffrontpage \bumpfootsymbolpos \else \advance\lastf@@t by 1
     \ifPhysRev \bumpfootsymbolneg \else \bumpfootsymbolpos \fi \fi
   \gl@bal\lastf@@t=\pagen@ }
\def\bumpfootsymbolpos{\ifnum\footsymbolcount <0
                            \gl@bal\footsymbolcount =0 \fi
    \ifnum\lastf@@t<\pagen@ \gl@bal\footsymbolcount=0
     \else \gl@bal\advance\footsymbolcount by 1 \fi }
\def\bumpfootsymbolneg{\ifnum\footsymbolcount >0
             \gl@bal\footsymbolcount =0 \fi
         \gl@bal\advance\footsymbolcount by -1 }
\def\fd@f#1 {\xdef\footsymbol{\mathchar"#1 }}
\def\generatefootsymbol{\ifcase\footsymbolcount \fd@f 13F \or \fd@f 279
        \or \fd@f 27A \or \fd@f 278 \or \fd@f 27B \else
        \ifnum\footsymbolcount <0 \fd@f{023 \number-\footsymbolcount }
         \else \fd@f 203 {\loop \ifnum\footsymbolcount >5
                \fd@f{203 \footsymbol } \advance\footsymbolcount by -1
                \repeat }\fi \fi }

\def\nonfrenchspacing{\sfcode`\.=3001 \sfcode`\!=3000 \sfcode`\?=3000
        \sfcode`\:=2000 \sfcode`\;=1500 \sfcode`\,=1251 }
\nonfrenchspacing
\newdimen\d@twidth
{\setbox0=\hbox{s.} \gl@bal\d@twidth=\wd0 \setbox0=\hbox{s}
        \gl@bal\advance\d@twidth by -\wd0 }
\def\removehglue{\loop \unskip \ifdim\lastskip >\z@ \repeat }
\def\roll@ver#1{\removehglue \nobreak \count255 =\spacefactor \dimen@=\z@
        \ifnum\count255 =3001 \dimen@=\d@twidth \fi
        \ifnum\count255 =1251 \dimen@=\d@twidth \fi
    \iftwelv@ \kern-\dimen@ \else \kern-0.83\dimen@ \fi
   #1\spacefactor=\count255 }
\def\step@ver#1{\rel@x \ifmmode #1\else \ifhmode
        \roll@ver{${}#1$}\else {\setbox0=\hbox{${}#1$}}\fi\fi }
\def\attach#1{\step@ver{\strut^{\mkern 2mu #1} }}
%
%
%
\newcount\chapternumber      \chapternumber=0
\newcount\sectionnumber      \sectionnumber=0
\newcount\equanumber         \equanumber=0
\let\chapterlabel=\rel@x
\let\sectionlabel=\rel@x
\newtoks\chapterstyle        \chapterstyle={\Number}
\newtoks\sectionstyle        \sectionstyle={\Number}
\newskip\chapterskip         \chapterskip=\bigskipamount
\newskip\sectionskip         \sectionskip=\medskipamount
\newskip\headskip            \headskip=8pt plus 3pt minus 3pt
\newdimen\chapterminspace    \chapterminspace=15pc
\newdimen\sectionminspace    \sectionminspace=10pc
\newdimen\referenceminspace  \referenceminspace=20pc
\newif\ifcn@                 \cn@true
\newif\ifcn@@                \cn@@false
\def\numberedchapters{\cn@true}
\def\unnumberedchapters{\cn@false\sequentialequations}
\def\chapterreset{\gl@bal\advance\chapternumber by 1
   \ifnum\equanumber<0 \else\gl@bal\equanumber=0\fi
   \sectionnumber=0 \let\sectionlabel=\rel@x
   \ifcn@ \gl@bal\cn@@true {\pr@tect
       \xdef\chapterlabel{\the\chapterstyle{\the\chapternumber}}}%
    \else \gl@bal\cn@@false \gdef\chapterlabel{\rel@x}\fi }
\def\@alpha#1{\count255='140 \advance\count255 by #1\char\count255}
 \def\alphabetic{\n@expand\@alpha}
\def\@Alpha#1{\count255='100 \advance\count255 by #1\char\count255}
 \def\Alphabetic{\n@expand\@Alpha}
\def\@Roman#1{\uppercase\expandafter{\romannumeral #1}}
 \def\Roman{\n@expand\@Roman}
\def\@roman#1{\romannumeral #1}    \def\roman{\n@expand\@roman}
\def\@number#1{\number #1}         \def\Number{\n@expand\@number}
\def\BLANK#1{\rel@x}               
\def\titleparagraphs{\interlinepenalty=9999
     \leftskip=0.03\hsize plus 0.22\hsize minus 0.03\hsize
     \rightskip=\leftskip \parfillskip=0pt
     \hyphenpenalty=9000 \exhyphenpenalty=9000
     \tolerance=9999 \pretolerance=9000
     \spaceskip=0.333em \xspaceskip=0.5em }
\def\titlestyle#1{\par\begingroup \titleparagraphs
     \iftwelv@\fourteenpoint\else\twelvepoint\fi
   \noindent #1\par\endgroup }
\def\spacecheck#1{\dimen@=\pagegoal\advance\dimen@ by -\pagetotal
   \ifdim\dimen@<#1 \ifdim\dimen@>0pt \vfil\break \fi\fi}
\def\chapter#1{\par \penalty-300 \vskip\chapterskip
   \spacecheck\chapterminspace
   \chapterreset \titlestyle{\ifcn@@\chapterlabel.~\fi #1}
   \nobreak\vskip\headskip \penalty 30000
   {\pr@tect\wlog{\string\chapter\space \chapterlabel}} }

\def\section#1{\par \ifnum\lastpenalty=30000\else
   \penalty-200\vskip\sectionskip \spacecheck\sectionminspace\fi
   \gl@bal\advance\sectionnumber by 1
   {\pr@tect
   \xdef\sectionlabel{\ifcn@@ \chapterlabel.\fi
       \the\sectionstyle{\the\sectionnumber}}%
   \wlog{\string\section\space \sectionlabel}}%
   \noindent {\caps\enspace\sectionlabel.~~#1}\par
   \nobreak\vskip\headskip \penalty 30000 }
\def\subsection#1{\par
   \ifnum\the\lastpenalty=30000\else \penalty-100\smallskip \fi
   \noindent\undertext{#1}\enspace \vadjust{\penalty5000}}

\def\undertext#1{\vtop{\hbox{#1}\kern 1pt \hrule}}
\def\ACK{\par\penalty-100\medskip \spacecheck\sectionminspace
   \line{\fourteenrm\hfil ACKNOWLEDGEMENTS\hfil}\nobreak\vskip\headskip }

\def\APPENDIX#1#2{\par\penalty-300\vskip\chapterskip
   \spacecheck\chapterminspace \chapterreset \xdef\chapterlabel{#1}
   \titlestyle{APPENDIX #2} \nobreak\vskip\headskip \penalty 30000
   \wlog{\string\Appendix~\chapterlabel} }
\def\Appendix#1{\APPENDIX{#1}{#1}}
\def\appendix{\APPENDIX{A}{}}
%
%
%
\def\eqname#1{\rel@x {\pr@tect
  \ifnum\equanumber<0 \xdef#1{{\rm(\number-\equanumber)}}%
     \gl@bal\advance\equanumber by -1
  \else \gl@bal\advance\equanumber by 1
   \xdef#1{{\rm(\ifcn@@ \chapterlabel.\fi \number\equanumber)}}\fi
  }#1}

\def\eqn{\eqno\eqname}

\def\eqinsert#1{\noalign{\dimen@=\prevdepth \nointerlineskip
   \setbox0=\hbox to\displaywidth{\hfil #1}
   \vbox to 0pt{\kern 0.5\baselineskip\hbox{$\!\box0\!$}\vss}
   \prevdepth=\dimen@}}
%

%
%
\def\GENITEM#1;#2{\par \hangafter=0 \hangindent=#1
    \Textindent{$ #2 $}\ignorespaces}
\outer\def\newitem#1=#2;{\gdef#1{\GENITEM #2;}}

\newdimen\itemsize                \itemsize=30pt
\newitem\item=1\itemsize;
\newitem\sitem=1.75\itemsize;     
\newitem\ssitem=2.5\itemsize;     
\outer\def\newlist#1=#2&#3&#4;{\toks0={#2}\toks1={#3}%
   \count255=\escapechar \escapechar=-1
   \alloc@0\list\countdef\insc@unt\listcount     \listcount=0
   \edef#1{\par
      \countdef\listcount=\the\allocationnumber
      \advance\listcount by 1
      \hangafter=0 \hangindent=#4
      \Textindent{\the\toks0{\listcount}\the\toks1}}
   \expandafter\expandafter\expandafter
    \edef\c@t#1{begin}{\par
      \countdef\listcount=\the\allocationnumber \listcount=1
      \hangafter=0 \hangindent=#4
      \Textindent{\the\toks0{\listcount}\the\toks1}}
   \expandafter\expandafter\expandafter
    \edef\c@t#1{con}{\par \hangafter=0 \hangindent=#4 \noindent}
   \escapechar=\count255}
\def\c@t#1#2{\csname\string#1#2\endcsname}
\newlist\point=\Number&.&1.0\itemsize;
\newlist\subpoint=(\alphabetic&)&1.75\itemsize;
\newlist\subsubpoint=(\roman&)&2.5\itemsize;
%

%
%
%
%
\newcount\referencecount     \referencecount=0
\newcount\lastrefsbegincount \lastrefsbegincount=0
\newif\ifreferenceopen       \newwrite\referencewrite
\newdimen\refindent          \refindent=30pt
\def\normalrefmark#1{\attach{\scriptscriptstyle [ #1 ] }}
\let\PRrefmark=\attach
\def\NPrefmark#1{\step@ver{{\;[#1]}}}
\def\refmark#1{\rel@x\ifPhysRev\PRrefmark{#1}\else\normalrefmark{#1}\fi}
\def\refend@{\refmark{\number\referencecount}}
\def\refend{\refend@{}\space }
\def\refsend{\refmark{\count255=\referencecount
   \advance\count255 by-\lastrefsbegincount
   \ifcase\count255 \number\referencecount
   \or \number\lastrefsbegincount,\number\referencecount
   \else \number\lastrefsbegincount-\number\referencecount \fi}\space }
\def\REFNUM#1{\rel@x \gl@bal\advance\referencecount by 1
    \xdef#1{\the\referencecount }}
\def\Refnum#1{\REFNUM #1\refend@ } 
\def\REF#1{\REFNUM #1\R@FWRITE\ignorespaces}
\def\Ref#1{\Refnum #1\REFWRITE }
\def\ref{\Ref\?}
\def\REFS#1{\REFNUM #1\gl@bal\lastrefsbegincount=\referencecount
    \REFWRITE }

       \let\REFSCON=\REF
\def\r@fitem#1{\par \hangafter=0 \hangindent=\refindent \Textindent{#1}}
\def\refitem#1{\r@fitem{#1.}}
\def\NPrefitem#1{\r@fitem{[#1]}}
\def\NPrefs{\let\refmark=\NPrefmark \let\refitem=NPrefitem}
\def\REFWRITE{\R@FWRITE\rel@x }
\def\R@FWRITE#1{\ifreferenceopen \else \gl@bal\referenceopentrue
     \immediate\openout\referencewrite=\jobname.refs
     \toks@={\begingroup \refoutspecials \catcode`\^^M=10 }%
     \immediate\write\referencewrite{\the\toks@}\fi
    \immediate\write\referencewrite{\noexpand\refitem %
                                    {\the\referencecount}}%
    \p@rse@ndwrite \referencewrite #1}
\begingroup
 \catcode`\^^M=\active \let^^M=\relax %
 \gdef\p@rse@ndwrite#1#2{\begingroup \catcode`\^^M=12 \newlinechar=`\^^M%
         \chardef\rw@write=#1\sc@nlines#2}%
 \gdef\sc@nlines#1#2{\sc@n@line \g@rbage #2^^M\endsc@n \endgroup #1}%
 \gdef\sc@n@line#1^^M{\expandafter\toks@\expandafter{\deg@rbage #1}%
         \immediate\write\rw@write{\the\toks@}%
         \futurelet\n@xt \sc@ntest }%
\endgroup
\def\sc@ntest{\ifx\n@xt\endsc@n \let\n@xt=\rel@x
       \else \let\n@xt=\sc@n@notherline \fi \n@xt }
\def\sc@n@notherline{\sc@n@line \g@rbage }
\def\deg@rbage#1{}
\let\g@rbage=\relax    \let\endsc@n=\relax
\def\refout{\par\penalty-400\vskip\chapterskip
   \spacecheck\referenceminspace
   \ifreferenceopen \Closeout\referencewrite \referenceopenfalse \fi
   \line{\fourteenrm\hfil REFERENCES\hfil}\vskip\headskip
   \input \jobname.refs
   }
\def\refoutspecials{\sfcode`\.=1000 \interlinepenalty=1000
         \rightskip=\z@ plus 1em minus \z@ }
\def\Closeout#1{\toks0={\par\endgroup}\immediate\write#1{\the\toks0}%
   \immediate\closeout#1}
%
%
\newcount\figurecount     \figurecount=0
\newcount\tablecount      \tablecount=0
\newif\iffigureopen       \newwrite\figurewrite
\newif\iftableopen        \newwrite\tablewrite
\def\FIGNUM#1{\rel@x \gl@bal\advance\figurecount by 1
    \xdef#1{\the\figurecount}}
\def\FIGURE#1{\FIGNUM #1\F@GWRITE\ignorespaces }

\def\figitem#1{\r@fitem{#1)}}
\def\FIGWRITE{\F@GWRITE\rel@x }
\def\TABNUM#1{\rel@x \gl@bal\advance\tablecount by 1
    \xdef#1{\the\tablecount}}
\def\TABLE#1{\TABNUM #1\T@BWRITE\ignorespaces }

\def\tabitem#1{\r@fitem{#1:}}
\def\TABWRITE{\T@BWRITE\rel@x }
\def\F@GWRITE#1{\iffigureopen \else \gl@bal\figureopentrue
     \immediate\openout\figurewrite=\jobname.figs
     \toks@={\begingroup \catcode`\^^M=10 }%
     \immediate\write\figurewrite{\the\toks@}\fi
    \immediate\write\figurewrite{\noexpand\figitem %
                                 {\the\figurecount}}%
    \p@rse@ndwrite \figurewrite #1}
\def\T@BWRITE#1{\iftableopen \else \gl@bal\tableopentrue
     \immediate\openout\tablewrite=\jobname.tabs
     \toks@={\begingroup \catcode`\^^M=10 }%
     \immediate\write\tablewrite{\the\toks@}\fi
    \immediate\write\tablewrite{\noexpand\tabitem %
                                 {\the\tablecount}}%
    \p@rse@ndwrite \tablewrite #1}
\def\figout{\par\penalty-400
   \vskip\chapterskip\spacecheck\referenceminspace
   \iffigureopen \Closeout\figurewrite \figureopenfalse \fi
   \line{\fourteenrm\hfil FIGURE CAPTIONS\hfil}\vskip\headskip
   \input \jobname.figs
   }
\def\tabout{\par\penalty-400
   \vskip\chapterskip\spacecheck\referenceminspace
   \iftableopen \Closeout\tablewrite \tableopenfalse \fi
   \line{\fourteenrm\hfil TABLE CAPTIONS\hfil}\vskip\headskip
   \input \jobname.tabs
   }
%
%
%
\newbox\picturebox
\def\p@cht{\ht\picturebox }
\def\p@cwd{\wd\picturebox }
\def\p@cdp{\dp\picturebox }
\newdimen\xshift
\newdimen\yshift
\newdimen\captionwidth
\newskip\captionskip
\captionskip=15pt plus 5pt minus 3pt
\def\fullwidth{\captionwidth=\hsize }
\newtoks\Caption
\newif\ifcaptioned
\newif\ifselfcaptioned
\def\caption{\captionedtrue \Caption }
\newcount\linesabove
\newif\iffileexists
\newtoks\picfilename
\def\fil@#1 {\fileexiststrue \picfilename={#1}}
\def\file#1{\if=#1\let\n@xt=\fil@ \else \def\n@xt{\fil@ #1}\fi \n@xt }
\def\pl@t{\begingroup \pr@tect
    \setbox\picturebox=\hbox{}\fileexistsfalse
    \let\height=\p@cht \let\width=\p@cwd \let\depth=\p@cdp
    \xshift=\z@ \yshift=\z@ \captionwidth=\z@
    \Caption={}\captionedfalse
    \linesabove =0 \picturedefault }
\def\plot{\pl@t \selfcaptionedfalse }
\def\Picture#1{\gl@bal\advance\figurecount by 1
    \xdef#1{\the\figurecount}\pl@t \selfcaptionedtrue }

\def\s@vepicture{\iffileexists \parsefilename \redopicturebox \fi
   \ifdim\captionwidth>\z@ \else \captionwidth=\p@cwd \fi
   \xdef\lastpicture{\iffileexists
        \setbox0=\hbox{\raise\the\yshift \vbox{%
              \moveright\the\xshift\hbox{\picturedefinition}}}%
        \else \setbox0=\hbox{}\fi
         \ht0=\the\p@cht \wd0=\the\p@cwd \dp0=\the\p@cdp
         \vbox{\hsize=\the\captionwidth \line{\hss\box0 \hss }%
              \ifcaptioned \vskip\the\captionskip \noexpand\Tenpoint
                \ifselfcaptioned Figure~\the\figurecount.\enspace \fi
                \the\Caption \fi }}%
    \endgroup }
\let\endpicture=\s@vepicture
\def\savepicture#1{\s@vepicture \global\let#1=\lastpicture }
\def\displaypicture{\fullwidth \s@vepicture $$\lastpicture $${}}
\def\toppicture{\fullwidth \s@vepicture \topinsert
    \lastpicture \medskip \endinsert }
\def\midpicture{\fullwidth \s@vepicture \midinsert
    \lastpicture \endinsert }
%
%
\def\leftpicture{\pres@tpicture
    \dimen@i=\hsize \advance\dimen@i by -\dimen@ii
    \setbox\picturebox=\hbox to \hsize {\box0 \hss }%
    \wr@paround }
\def\rightpicture{\pres@tpicture
    \dimen@i=\z@
    \setbox\picturebox=\hbox to \hsize {\hss \box0 }%
    \wr@paround }
\def\pres@tpicture{\gl@bal\linesabove=\linesabove
    \s@vepicture \setbox\picturebox=\vbox{
         \kern \linesabove\baselineskip \kern 0.3\baselineskip
         \lastpicture \kern 0.3\baselineskip }%
    \dimen@=\p@cht \dimen@i=\dimen@
    \advance\dimen@i by \pagetotal
    \par \ifdim\dimen@i>\pagegoal \vfil\break \fi
    \dimen@ii=\hsize
    \advance\dimen@ii by -\parindent \advance\dimen@ii by -\p@cwd
    \setbox0=\vbox to\z@{\kern-\baselineskip \unvbox\picturebox \vss }}
\def\wr@paround{\Caption={}\count255=1
    \loop \ifnum \linesabove >0
         \advance\linesabove by -1 \advance\count255 by 1
         \advance\dimen@ by -\baselineskip
         \expandafter\Caption \expandafter{\the\Caption \z@ \hsize }%
      \repeat
    \loop \ifdim \dimen@ >\z@
         \advance\count255 by 1 \advance\dimen@ by -\baselineskip
         \expandafter\Caption \expandafter{%
             \the\Caption \dimen@i \dimen@ii }%
      \repeat
    \edef\n@xt{\parshape=\the\count255 \the\Caption \z@ \hsize }%
    \par\noindent \n@xt \strut \vadjust{\box\picturebox }}
\let\picturedefault=\relax
\let\parsefilename=\relax
\def\redopicturebox{\let\picturedefinition=\rel@x
   \errhelp=\disabledpictures
   \errmessage{This version of TeX cannot handle pictures.  Sorry.}}
\newhelp\disabledpictures
     {You will get a blank box in place of your picture.}
%
%
%
%
%
%
%
%
%
%
\def\FRONTPAGE{\ifvoid255\else\vfill\penalty-20000\fi
   \gl@bal\pagenumber=1     \gl@bal\chapternumber=0
   \gl@bal\equanumber=0     \gl@bal\sectionnumber=0
   \gl@bal\referencecount=0 \gl@bal\figurecount=0
   \gl@bal\tablecount=0     \gl@bal\frontpagetrue
   \gl@bal\lastf@@t=0       \gl@bal\footsymbolcount=0
   \gl@bal\cn@@false }

\def\papers{\papersize\headline=\paperheadline\footline=\paperfootline}
\def\papersize{\hsize=35pc \vsize=50pc \hoffset=0pc \voffset=1pc
   \advance\hoffset by\HOFFSET \advance\voffset by\VOFFSET
   \pagebottomfiller=0pc
   \skip\footins=\bigskipamount \normalspace }
\papers  
%
%
\newskip\lettertopskip       \lettertopskip=20pt plus 50pt
\newskip\letterbottomskip    \letterbottomskip=\z@ plus 100pt
\newskip\signatureskip       \signatureskip=40pt plus 3pt
\def\lettersize{\hsize=6.5in \vsize=8.5in \hoffset=0in \voffset=0.5in
   \advance\hoffset by\HOFFSET \advance\voffset by\VOFFSET
   \pagebottomfiller=\letterbottomskip
   \skip\footins=\smallskipamount \multiply\skip\footins by 3
   \singlespace }
\def\MEMO{\lettersize \headline=\letterheadline \footline={\hfil }%
   \let\rule=\memorule \FRONTPAGE \memohead }

\def\memodate{\afterassignment\MEMO \date }
\def\memit@m#1{\smallskip \hangafter=0 \hangindent=1in
    \Textindent{\caps #1}}
\def\subject{\memit@m{Subject:}}
\def\topic{\memit@m{Topic:}}
\def\from{\memit@m{From:}}
\def\to{\rel@x \ifmmode \rightarrow \else \memit@m{To:}\fi }
\def\memorule{\medskip\hrule height 1pt\bigskip}  
\def\memohead{\centerline{\fourteenrm MEMORANDUM}}
\newwrite\labelswrite
\newtoks\rw@toks
\def\letters{\lettersize
   \headline=\letterheadline \footline=\letterfootline
   \immediate\openout\labelswrite=\jobname.lab}

\let\letterhead=\rel@x
\def\addressee#1{\medskip\line{\hskip 0.75\hsize plus\z@ minus 0.25\hsize
                               \the\date \hfil }%
   \vskip \lettertopskip
   \ialign to\hsize{\strut ##\hfil\tabskip 0pt plus \hsize \crcr #1\crcr}
   \writelabel{#1}\medskip \noindent\hskip -\spaceskip \ignorespaces }
\def\rwl@begin#1\cr{\rw@toks={#1\crcr}\rel@x
   \immediate\write\labelswrite{\the\rw@toks}\futurelet\n@xt\rwl@next}
\def\rwl@next{\ifx\n@xt\rwl@end \let\n@xt=\rel@x
      \else \let\n@xt=\rwl@begin \fi \n@xt}
\let\rwl@end=\rel@x
\def\writelabel#1{\immediate\write\labelswrite{\noexpand\labelbegin}
     \rwl@begin #1\cr\rwl@end
     \immediate\write\labelswrite{\noexpand\labelend}}
\newtoks\FromAddress         \FromAddress={}
\newtoks\sendername          \sendername={}
\newbox\FromLabelBox
\newdimen\labelwidth          \labelwidth=6in
\def\makelabels{\afterassignment\Makelabels \sendername=}
\def\Makelabels{\FRONTPAGE \letterinfo={\hfil } \MakeFromBox
     \immediate\closeout\labelswrite  \input \jobname.lab\vfil\eject}
\let\labelend=\rel@x
\def\labelbegin#1\labelend{\setbox0=\vbox{\ialign{##\hfil\cr #1\crcr}}
     \MakeALabel }
\def\MakeFromBox{\gl@bal\setbox\FromLabelBox=\vbox{\Tenpoint
     \ialign{##\hfil\cr \the\sendername \the\FromAddress \crcr }}}
\def\MakeALabel{\vskip 1pt \hbox{\vrule \vbox{
        \hsize=\labelwidth \hrule\bigskip
        \leftline{\hskip 1\parindent \copy\FromLabelBox}\bigskip
        \centerline{\hfil \box0 } \bigskip \hrule
        }\vrule } \vskip 1pt plus 1fil }
\def\signed#1{\par \nobreak \bigskip \dt@pfalse \begingroup
  \everycr={\noalign{\nobreak
            \ifdt@p\vskip\signatureskip\gl@bal\dt@pfalse\fi }}%
  \tabskip=0.5\hsize plus \z@ minus 0.5\hsize
  \halign to\hsize {\strut ##\hfil\tabskip=\z@ plus 1fil minus \z@\crcr
          \noalign{\gl@bal\dt@ptrue}#1\crcr }%
  \endgroup \bigskip }
\newbox\letterb@x
\def\lettertext{\par \vskip\parskip \unvcopy\letterb@x \par }
\def\multiletter{\setbox\letterb@x=\vbox\bgroup
      \everypar{\vrule height 1\baselineskip depth 0pt width 0pt }
      \singlespace \topskip=\baselineskip }
\def\letterend{\par\egroup}
%
%
%
\newskip\frontpageskip
\newtoks\Pubnum   
\newtoks\Pubtype  \let\pubtype=\Pubtype
\newif\ifp@bblock  \p@bblocktrue
\def\PH@SR@V{\doubl@true \baselineskip=24.1pt plus 0.2pt minus 0.1pt
             \parskip= 3pt plus 2pt minus 1pt }
\def\PHYSREV{\papers\PhysRevtrue\PH@SR@V}

\def\titlepage{\FRONTPAGE\papers\ifPhysRev\PH@SR@V\fi
   \ifp@bblock\p@bblock \else\hrule height\z@ \rel@x \fi }
\def\nopubblock{\p@bblockfalse}
\def\endpage{\vfil\break}
\frontpageskip=12pt plus .5fil minus 2pt
\Pubtype={}
\Pubnum={}
\def\p@bblock{\begingroup \tabskip=\hsize minus \hsize
   \baselineskip=1.5\ht\strutbox \topspace-2\baselineskip
   \halign to\hsize{\strut ##\hfil\tabskip=0pt\crcr
       \the\Pubnum\crcr\the\date\crcr\the\pubtype\crcr}\endgroup}
\def\title#1{\vskip\frontpageskip \titlestyle{#1} \vskip\headskip }
\def\author#1{\vskip\frontpageskip\titlestyle{\twelvecp #1}\nobreak}

\def\address#1{\par\kern 5pt\titlestyle{\twelvepoint\it #1}}
\def\andaddress{\par\kern 5pt \centerline{\sl and} \address}

\def\abstract{\par\dimen@=\prevdepth \hrule height\z@ \prevdepth=\dimen@
   \vskip\frontpageskip\centerline{\fourteenrm ABSTRACT}\vskip\headskip }
\def\abstnotitle{\par\dimen@=\prevdepth \hrule height\z@ \prevdepth=\dimen@ }

%
%
%
\def\ie{\hbox{\it i.e.}}       
\def\eg{\hbox{\it e.g.}}       
   
\def\\{\rel@x \ifmmode \backslash \else {\tt\char`\\}\fi }
\def\sequentialequations{\rel@x \if\equanumber<0 \else
  \gl@bal\equanumber=-\equanumber \gl@bal\advance\equanumber by -1 \fi }
\def\journal#1&#2(#3){\begingroup \let\journal=\dummyj@urnal
    \unskip, \sl #1\unskip~\bf\ignorespaces #2\rm
    (\afterassignment\j@ur \count255=#3), \endgroup\ignorespaces }
\def\j@ur{\ifnum\count255<100 \advance\count255 by 1900 \fi
          \number\count255 }
\def\dummyj@urnal{%
    \toks@={Reference foul up: nested \journal macros}%
    \errhelp={Your forgot & or ( ) after the last \journal}%
    \errmessage{\the\toks@ }}
\def\cropen#1{\crcr\noalign{\vskip #1}}
\def\crr{\cropen{3\jot }}
\def\topspace{\hrule height 0pt depth 0pt \vskip}

\def\half{\coeff12 }

\def\Buildrel#1\under#2{\mathrel{\mathop{#2}\limits_{#1}}}
\def\becomes#1{\mathchoice{\becomes@\scriptstyle{#1}}
   {\becomes@\scriptstyle{#1}} {\becomes@\scriptscriptstyle{#1}}
   {\becomes@\scriptscriptstyle{#1}}}
\def\becomes@#1#2{\mathrel{\setbox0=\hbox{$\m@th #1{\,#2\,}$}%
        \mathop{\hbox to \wd0 {\rightarrowfill}}\limits_{#2}}}

\let\int=\intop         
\def\lsim{\mathrel{\mathpalette\@versim<}}
\def\gsim{\mathrel{\mathpalette\@versim>}}
\def\@versim#1#2{\vcenter{\offinterlineskip
        \ialign{$\m@th#1\hfil##\hfil$\crcr#2\crcr\sim\crcr } }}
\def\big#1{{\hbox{$\left#1\vbox to 0.85\b@gheight{}\right.\n@space$}}}
\def\Big#1{{\hbox{$\left#1\vbox to 1.15\b@gheight{}\right.\n@space$}}}
\def\bigg#1{{\hbox{$\left#1\vbox to 1.45\b@gheight{}\right.\n@space$}}}
\def\Bigg#1{{\hbox{$\left#1\vbox to 1.75\b@gheight{}\right.\n@space$}}}
\def\){\mskip 2mu\nobreak }
%
%
%
\let\sec@nt=\sec
\def\sec{\rel@x\ifmmode\let\n@xt=\sec@nt\else\let\n@xt\section\fi\n@xt}
\def\obsolete#1{\message{Macro \string #1 is obsolete.}}
\def\firstsec#1{\obsolete\firstsec \section{#1}}
\def\firstsubsec#1{\obsolete\firstsubsec \subsection{#1}}
\def\thispage#1{\obsolete\thispage \gl@bal\pagenumber=#1\frontpagefalse}
\def\thischapter#1{\obsolete\thischapter \gl@bal\chapternumber=#1}
\def\splitout{\obsolete\splitout\rel@x}
\def\prop{\obsolete\prop \propto }
\def\nextequation#1{\obsolete\nextequation \gl@bal\equanumber=#1
   \ifnum\the\equanumber>0 \gl@bal\advance\equanumber by 1 \fi}
\def\BOXITEM{\afterassigment\B@XITEM\setbox0=}
\def\B@XITEM{\par\hangindent\wd0 \noindent\box0 }
%
%
%
\def\phyzzx{PHY\setbox0=\hbox{Z}\copy0 \kern-0.5\wd0 \box0 X}
        
\message{ by V.K.}
%

%
%
%
\def\slacpub{\afterassignment\slacp@b\toks@}
\def\slacp@b{\edef\n@xt{\Pubnum={SLAC--PUB--\the\toks@}}\n@xt}

\expandafter\ifx\csname eightrm\endcsname\relax
    \let\eightrm=\ninerm \let\eightbf=\ninebf \fi
\def\memohead{\hrule height\z@ \kern -0.5in
    \line{\quad\fourteenrm SLAC MEMORANDUM\hfil \twelverm\the\date\quad}}
\def\memorule{\par \medskip \hrule height 0.5pt \kern 1.5pt
   \hrule height 0.5pt \medskip}
\def\SLACHEAD{\setbox0=\vtop{\baselineskip=10pt
     \ialign{\eightrm ##\hfil\cr
        Theory Group, LNS\cr
        Cornell University\cr
        Ithaca, NY 14853-5001\cropen{1\jot}
        \cornphone\cr }}%
   \setbox2=\hbox{\caps Newman Laboratory of Nuclear Studies}%
   \hrule height \z@ \kern -0.5in
   \vbox to 0pt{\vss\centerline{\seventeenrm CORNELL UNIVERSITY}}
   \vbox{} \medskip
   \line{\hbox to 0.7\hsize{\hss \lower 10pt \box2 \hfill }\hfil
         \hbox to 0.25\hsize{\box0 \hfil }}\medskip }

\def\headindent{-.75 truein}
\font\bigbold=cmr10 at 17.3truept
\font\bigss=cmss10 at 12truept
\font\smallss=cmss9 at 9truept
\font\smallheadfont=cmr9 at 9truept
\def\today{\ifcase\month\or
  January\or February\or March\or April\or May\or June\or
  July\or August\or September\or October\or November\or December\fi
  \space\number\day, \number\year}
\raggedbottom
\parindent=0pt\parskip=0pt\nopagenumbers
\newdimen\longindentation \longindentation=4truein
\advance\longindentation by-\hoffset

\def\up#1{\leavevmode \raise.16ex\hbox{#1}}
\def\sendingaddress{{\smallheadfont
\line{\hskip .75truein \hskip \headindent
          Keith R. Dienes \hfill {\smallss Bitnet:} dien@crnlnuc}
\line{\hskip.75truein \hskip  \headindent
          (607) 255-5725 / {\smallss Secretary:} (607) 255-5722 \hfill
{\smallss Internet:} dien@lnssun1.tn.cornell.edu}
\line{\hskip.75truein \hskip  \headindent
         {\smallss  Home phone:}  (607) 272-6176 \hfill}
}}
\def\univletterhead{\advance\hsize by .5truein
\vglue-.4truein\vglue-\voffset%
\vskip 1in
\hskip-1in
\includegraphics{/mu/jjd/letters/crest.ps}
\vskip -1.2in
\line{{\hskip .75truein\hskip\headindent \bigbold Cornell University \hfill }}
\vskip 3truept
\line{{\hskip .75truein \hskip\headindent
\bigss Floyd R. Newman Laboratory of Nuclear Studies \hfill}}
\vskip -6truept
\line{\hskip .75truein\hskip\headindent \hrulefill}
\line{{\hskip .75truein
\hskip\headindent \smallss Newman Laboratory \hfill Telex: WUI 6713054}}
\vskip -3truept
\line{{\hskip .75truein\hskip \headindent
\smallss Ithaca, NY 14853-5001 \hfill Telefax: 607-254-4552}}
\sendingaddress\par
\vskip .5truein
\advance\hsize by -.5truein
}
\let\letterhead=\univletterhead
\FromAddress={\crcr \slacbin \cr
    P.\ O.\ Box 4349\cr Stanford, California 94309\cr }
\def\slacbin{SLAC\expandafter\ifx\csname binno\endcsname\relax
                             \else , Bin \binno \fi }

\def\cornphone{(607) 255--\cornext}

\def\cornext{5725}
\VOFFSET=33pt
\papersize
%
%
\newwrite\figscalewrite
\newif\iffigscaleopen
\newif\ifgrayscale
\newif\ifreadyfile
\def\picturedefault{\grayscalefalse \readyfilefalse
    \gdef\ready{\readyfiletrue}\gdef\gray{\ready\grayscaletrue}}
\def\parsefilename{\ifreadyfile \else
    \iffigscaleopen \else \gl@bal\figscaleopentrue
       \immediate\openout\figscalewrite=\jobname.scalecon \fi
    \toks0={ }\immediate\write\figscalewrite{%
       \the\p@cwd \the\toks0 \the\p@cht \the\toks0 \the\picfilename }%
    \expandafter\p@rse \the\picfilename..\endp@rse \fi }
\def\p@rse#1.#2.#3\endp@rse{%
   \if*#3*\dop@rse #1.1..\else \if.#3\dop@rse #1.1..\else
                                \dop@rse #1.#3\fi \fi
   \expandafter\picfilename\expandafter{\n@xt}}
\def\dop@rse#1.#2..{\count255=#2 \ifnum\count255<1 \count255=1 \fi
   \ifnum\count255<10  \edef\n@xt{#1.PICT00\the\count255}\else
   \ifnum\count255<100 \edef\n@xt{#1.PICT0\the\count255}\else
                       \edef\n@xt{#1.PICT\the\count255}\fi\fi }
\def\redopicturebox{\edef\picturedefinition{\ifgrayscale
     \special{insert(\the\picfilename)}\else
     \special{mergeug(\the\picfilename)}\fi }}
%
%

\let\rule=\memorule

\def\boxit#1{\vbox{\hrule\hbox{\vrule\kern3pt
\vbox{\kern3pt#1\kern3pt}\kern3pt\vrule}\hrule}}

\catcode`\@=12 
%

\catcode`\@=11
\def\Smallpoint{\rel@x
\def\Tinpoint{\tenpoint \rel@x
\ifsingl@\subspaces@t2:5;\else\subspaces@t3:5;\fi
\ifdoubl@ \multiply\baselineskip by 5
\divide\baselineskip by 4\fi}
\parindent=16.67pt
\itemsize=25pt
\thinmuskip=2.5mu
\medmuskip=3.33mu plus 1.67mu minus 3.33mu
\thickmuskip=4.17mu plus 4.17mu
\def\thinspace{\kern .13889em}
\def\negthinspace{\kern-.13889em}
\def\enspace{\kern.416667em}
\def\enskip{\hskip.416667em\rel@x}
\def\quad{\hskip.83333em\rel@x}
\def\qquad{\hskip1.66667em\rel@x}
\def\crr{\cropen{8.3333pt}}
\def\papersize{\vsize=39.0pc\hsize=28.6pc
\hoffset=-.55truein\voffset=.7pc
\skip\footins\bigskipamount}
\Tenpoint\papersize}
\catcode`\@=12
\newbox\leftpage
\newdimen\fullhsize
\newdimen\hstitle
\newdimen\hsbody
\def\LITTLE{\Smallpoint\relax
\let\lr=L
\hstitle=8truein\hsbody=4.75truein\fullhsize=10truein\hsize=\hsbody
\output={\ifnum\pageno=0
  \shipout\vbox{\hbox to \fullhsize{\hfill\pagebody\hfill}}\advancepageno
  \else
  \almostshipout{\leftline{\vbox{\pagebody\makefootline}}}\advancepageno
  \fi}
\def\almostshipout##1{\if L\lr
      \global\setbox\leftpage=##1 \global\let\lr=R
  \else
    \shipout\vbox{
      \hbox to\fullhsize{\box\leftpage\hfil##1}}  \global\let\lr=L\fi}
}


\pagenumbers

 ~
\vskip -0.22 truein
\date{CLNS 91/1113;  McGill/91-37\cr
hepth@xxx/9201078 \cr
January 1992 \cr}
\titlepage
\title{\bf NEW JACOBI-LIKE IDENTITIES FOR ${\bf Z}_K$ PARAFERMION CHARACTERS}
\medskip
\author{Philip C. Argyres$^1$\foot{E-mail address:
   pca@strange.tn.cornell.edu.}, Keith R. Dienes$^2$\foot{E-mail address:
   dien@hep.physics.mcgill.ca.},\break
 {\rm and~} S.-H. Henry Tye$^1$}
\smallskip
\address{$^1$ Newman Laboratory of Nuclear Studies \break
Cornell University \break
Ithaca, NY  14853-5001 ~~USA}
\smallskip
\address{$^2$ Dept.~of Physics, McGill University\break
E.~Rutherford Building, 3600 University St.\break
Montr\'eal, P.Q.~H3A-2T8 ~~Canada}
\bigskip
\abstract
We state and prove various new identities involving the
${\bf Z}_K$ parafermion characters (or level-$K$ string functions)
$c^\ell_n$ for the cases $K=4$, $K=8$, and $K=16$.  These identities
fall into three classes:  identities in the first
class are generalizations of the famous Jacobi $\vartheta$-function
identity (which is the $K=2$ special case), identities
in another class relate the level $K>2$ characters to
the Dedekind $\eta$-function, and identities in a third class relate
the $K>2$ characters to the Jacobi $\vartheta$-functions.
These identities play a crucial role in the interpretation of
fractional superstring spectra by indicating spacetime supersymmetry
and aiding in the identification of the spacetime spin and statistics
of fractional superstring states.
\endpage

\doublespace
\def\calZ{{\cal Z}}
\def\calF{{\cal F}}
\def\half{{\textstyle {1\over2}}}

\def\ie{{\it i.e.}}
\def\eg{{\it e.g.}}
\def\tautwo{{ \tau_2 }}
\def\thetatwo{{ \vartheta_2 }}
\def\thetathree{{ \vartheta_3 }}
\def\thetafour{{ \vartheta_4 }}

\def\deltaminushalf{{ \Delta^{-1/2} }}
\def\bZ{{\bf Z}}
\def\zk{{${\bf Z}_K$}}
\def\Gam{{\Gamma}}
\def\smallmatrix#1#2#3#4{{ {{#1}~{#2}\choose{#3}~{#4}} }}
\def\bone{{\bf 1}}
\def\equalmod#1{{\buildrel {#1} \over =}}

\hyphenation{pa-ra-fer-mion pa-ra-fer-mion-ic pa-ra-fer-mions }
\hyphenation{su-per-string frac-tion-ally su-per-re-pa-ra-met-ri-za-tion}
\hyphenation{su-per-sym-met-ric frac-tion-ally-su-per}
\hyphenation{space-time-super-sym-met-ric fer-mi-on}
\hyphenation{mod-u-lar mod-u-lar--in-var-i-ant}

\chapter{INTRODUCTION}

$\bZ_K$ parafermion theories\REF\Zam{
    A.B.~Zamolodchikov and V.A.~Fateev, {\it Sov.~Phys.~J.E.T.P.}
    {\bf 62} (1985) 215.}\refend\
have recently found a new application as the basic worldsheet building
blocks of fractional superstrings.\REF\AT{
   P.C.~Argyres and S.-H.H.~Tye, {\it Phys.~Rev.~Lett.}~{\bf 67} (1991)
   3339.}\refend\
Fractional superstrings are generalizations of the traditional
superstring and heterotic string, and are constructed essentially by replacing
the worldsheet supersymmetry of the superstring with a {\it fractional}\/
supersymmetry (parametrized by an integer $K$) which relates worldsheet
bosons not to fermions but to $\bZ_K$ {\it para}\/fermions.  It is
found that the critical spacetime dimensions of such string theories are less
than ten, and are in fact given by the simple formula
$$   D_c ~=~ 2 ~+~ {{16}\over K} ~,~~~~~K \geq 2 ~.
\eqn\kdimformula
$$
The special case $K=2$ reproduces the usual superstring and heterotic
string with critical dimension $D_c=10$, and the cases
with $K=4$, $K=8$, and $K=16$ yield new fractional superstring
theories with critical spacetime dimensions $D_c=6$, $D_c=4$, and $D_c=3$
respectively.

The worldsheet field content of these string theories consists in general
of bosons and $\bZ_K$ parafermions;  the special-case $\bZ_2$ parafermions
are equivalent to ordinary Majorana-Weyl fermions.  The partition
functions of fractional superstrings therefore involve the $\bZ_K$ parafermion
characters for $K>2$, just as the partition functions of the superstring
and heterotic string involve the ordinary ``$\bZ_2$'' characters of the
Majorana-Weyl fermions.  Similarly, just as the usual fermion characters
can be written in terms of the classical Jacobi theta-functions $\vartheta_i$,
the more general $\bZ_K$ parafermion characters can be written in terms of
the so-called ``string functions'' $c^\ell_n$ originally introduced
in the study of infinite-dimensional Lie algebras.\REF\Kacbook{
    See, \eg, V.G.~Ka\v{c}, {\it Infinite Dimensional
    Lie Algebras}, $3^{\rm rd}$ Edition (Cambridge University Press,
    1990).}\refend\
For $K=2$ these string functions are equivalent
to the Jacobi $\vartheta$-functions, but such is of course not the case
for $K>2$.

As is well-known in the $K=2$ superstring theory, the spacetime properties
of the string spectrum are reflected at the partition-function level
in the properties of these fermion characters.  For example,
any superstring or heterotic string spectrum exhibiting
a spacetime supersymmetry gives rise to a partition function
proportional to the factor
$$   J ~\equiv ~\thetathree^4\,-\, \thetatwo^4 \,-\,\thetafour^4~,
\eqn\kJdef
$$
and the well-known Jacobi identity $J=0$ is therefore responsible for the
vanishing of such partition functions at all mass levels of the theory.
This reflects the exact cancellation
of spacetime bosonic states (which arise from the
Neveu-Schwarz sector of the theory and yield
the terms $\thetathree^4 -\thetafour^4$ within $J$)
and spacetime fermionic states (which arise from the Ramond sector
and yield the term $\thetatwo^4$).

A similar situation exists for the $K>2$ fractional superstrings,
where once again we expect the spacetime properties of the fractional
superstring spectrum to be reflected in the partition function through
identities satisfied by the $\bZ_K$ parafermion characters (\ie, by
the string functions $c^\ell_n$).
Indeed, since the consistency of fractional superstrings
remains to be verified, the existence and use of
such identities provides an important step towards that goal.
In this paper we obtain and prove these new string-function
identities, and discuss as well their relevance to fractional superstrings.

In particular, we state and prove three series of identities.
First is a series of identities for the $\bZ_K$ parafermion
characters which are analogous to the $K=2$ Jacobi identity
and which can be considered to be its $K>2$ generalizations.
We will see that these new identities arise naturally in the partition
functions of fractional superstrings, and their
presence can therefore be interpreted as a signature
of spacetime supersymmetry in fractional superstring spectra.
For each value of $K$, we will find that the corresponding identity
in this series involves a vanishing combination of
only the $\bZ_K$ parafermion characters;
we will therefore, for reasons to become clear, refer to this series of
identities as the $[K,0]$ series.
Second, just as the individual terms within the $K=2$ Jacobi factor $J$
can be recognized as arising from either spacetime bosonic or fermionic
sectors,
we will see that a similar self-consistent grouping of terms is possible within
each of our new Jacobi-like identities for $K>2$.  This will result
in a second series of new string-function identities, each of which
(as we will see) relates the $K>2$ characters to the $K=2$ characters.
We will therefore refer to this series of identities as the $[K,2]$ series.
Finally, we find a third series of identities
which generalize another well-known $\vartheta$-function identity
$\thetatwo\thetathree\thetafour=2\eta^3$ (where $\eta$ is the Dedekind
$\eta$-function, the inverse of the boson character).  Just as this
identity relates the boson character $\eta^{-1}$ to the fermion
characters $\vartheta_i$, our third series of identities relates the
boson character $\eta^{-1}$ to the $\bZ_K$ {\it parafermion}\/ characters
$c^\ell_n$.  We will therefore refer to this series of identities as
the $[K,1]$ series, since (as we will see) the $K=1$ character is equal to
the boson character $\eta^{-1}$.  As our notation suggests,
these are undoubtedly only three of many such $[K_1,K_2]$-type series
of identities relating the characters of $\bZ_{K_1}$ and $\bZ_{K_2}$
parafermions to each other, and in this paper we also discuss how
such general $[K_1,K_2]$ identities may be obtained and proven.

Clearly, many of these new identities may have
interpretations in the theory of $\bZ_K$ parafermions
which are independent of fractional superstrings.
As a striking example, our $[16,2]$ identities indicate that
the simple algebraic {\it differences}\/ of many of the $\bZ_{16}$ characters
are nothing but the $\bZ_2$ (fermion) characters;  the implications
of this fact for the relationship between ordinary fermions and $\bZ_{16}$
parafermions are yet to be investigated.
Similar relationships exist as well for other values of $K$.
Therefore, we have organized this paper in such a way that
it can be read without a detailed understanding of fractional
superstrings {\it per se}.  In Sect.~2.1 we provide
a brief introduction to the $\bZ_K$ parafermion theory and
review the definitions and properties of the $\bZ_K$ parafermion
characters (or level-$K$ string functions).
In Sect.~2.2 we then discuss the role of the $\bZ_K$ parafermion
theory in the fractional superstring,
along the way introducing some of our new identities and discussing
the fractional-superstring contexts in which they arise.
Sect.~3 is simply a list of all of these new identities,
and in Sect.~4 we prove these identities using some powerful
results from the theory of modular functions.
The proofs of the $K>2$ identities exactly mirror the proofs of the
well-known $K=2$ special cases, and accordingly we have kept the discussion
in Sect.~4 sufficiently general so that it is clear how additional
$[K_1,K_2]$ identities may be obtained and proven.
A reader unconcerned with fractional superstrings can skip Sect.~2.2,
but it seems that the fractional superstring framework provides
interesting physical interpretations for many of these new identities.

\endpage

\chapter{$\bZ_K$ PARAFERMIONS AND FRACTIONAL SUPERSTRINGS}

In this section we first review the definitions and properties of
the $\bZ_K$ parafermion characters (or level-$K$ string functions)
which appear in our new identities.
We then provide a brief but self-contained introduction to the fractional
superstring idea, presenting many of these new identities
in the physical contexts in which they arise and discussing some
of their implications.

\section{The $\bZ_K$ Parafermion Characters}

The \zk\ parafermion theory\REF\Gep{D.~Gepner and Z.~Qiu,
    {\it Nucl.~Phys.} {\bf B285} (1987) 423.}\refmark{\Zam,\Gep}\
is closely related to, and in fact can be derived from, the
$SU(2)_K$ Wess-Zumino-Witten (WZW) theory.\REF\Knizh{V.G.~Knizhnik
    and A.B.~Zamolodchikov, {\it Nucl.~Phys.} {\bf B247} (1984) 83.}\refend\
As is well-known, the $SU(2)_K$ WZW theory can be viewed as a tensor
product of two independent theories:  the first is that of a free $U(1)$ boson
compactified on a circle of radius $2\sqrt{K}$, and the
remaining $SU(2)_K/U(1)$
coset theory is the \zk\ parafermion theory.  It therefore follows that
the characters of these remaining \zk\ parafermion fields can be obtained
from the full $SU(2)_K$ characters by appropriately factoring out the
the $U(1)$ boson characters.
We now review precisely how this is done.

We begin by considering the chiral $SU(2)_K$ WZW theory.\refmark{\Knizh}\
This theory consists of holomorphic primary fields $\Phi^j_m(z)$
which can be organized into $SU(2)$ representations labelled
by $j\in \bZ/2$, where $0\leq j\leq K/2$ and
$|m|\leq j$ with $j-m\in\bZ$.  Since $SU(2)_K$ always
has a $U(1)$ subgroup which can be bosonized as a free
boson $\varphi(z)$ compactified on a circle of radius $2\sqrt{K}$,
we can correspondingly factor the primary fields as
$$  \Phi^j_m(z) ~=~ \phi^j_m(z) \,
    \exp\left\lbrace {m\over {\sqrt{K}}}\,i\varphi(z)
    \right\rbrace ~.
\eqn\primfac
$$
The $\phi^j_m(z)$  are therefore primary fields of the
coset $SU(2)_K/U(1)$ theory,
\ie, of the $\bZ_K$ parafermion theory.  The parafermion fields
$\phi^j_m$ have conformal dimensions $h^{2j}_{2m}$ where
$$  h^\ell_n ~=~ {{\ell(\ell+2)}\over{4(K+2)}} ~-~ {{n^2}\over {4K}}
    \quad\quad\quad{\rm for}~~|n|\leq \ell~,
\eqn\pfdims
$$
and it is convenient to extend the definition of the parafermion
fields outside the range $|m|\leq j$ via the identifications
$$  \phi^j_m~=~\phi^j_{m+K}~=~\phi^{K/2-j}_{-(K/2-m)}~.
\eqn\phids
$$
The fusion rules of these parafermion fields $\phi^j_m$ follow
from those of the $SU(2)_K$ theory:
$$  [\phi^{j_1}_{m_1} ] \,\times \,
    [\phi^{j_2}_{m_2} ] ~=~ \sum_{j=|j_1-j_2|}^r \,
    [\phi^j_{m_1+m_2}]
\eqn\pfusion
$$
where $r={\rm min}(j_1+j_2,K-j_1-j_2)$ and where the sectors
$[\phi^j_m]$ include the primary fields $\phi^j_m$ and their
parafermion descendants.  Note that a given field $\phi^j_m$
may appear multiple times in the theory.

The $SU(2)_K$ characters $\chi_\ell(\tau,z)$ for spin $j=\ell/2$
are given in Ref.~[\Kacbook]:
$$  \chi_\ell(\tau,z)~=~{{\Theta_{\ell+1,K+2}(\tau,z)-
    \Theta_{-\ell-1,K+2}(\tau,z)}\over{\Theta_{1,2}(\tau,z)-
    \Theta_{-1,2}(\tau,z)}}~
\eqn\pSUtchars
$$
where the classical (Jacobi-Riemann) $\Theta$-functions are
defined by
$$  \Theta_{n,L}(\tau,z)~\equiv~\sum_{s\in\bZ+n/2L}
    \exp\left\lbrace 2\pi iL(s^2\tau-sz)\right\rbrace\quad\quad{\rm for}~
    n\in\bZ~~({\rm mod}~2L)~.
\eqn\pJRtheta
$$
These $\Theta$-functions have relatively simple
properties under modular transformations
$$    \tau ~\to~ \gamma\tau~\equiv~{{a\tau+b}\over{c\tau+d}}~,
\eqn\pmodact
$$
where $\gamma \equiv {a~b\choose c~d}$ is an element of the homogeneous
modular group $\tilde{\Gamma}\equiv
SL(2,\bZ)$ if $a,b,c,d\in\bZ$ and $ad-bc=1$.
The modular group $\Gamma\equiv \tilde{\Gamma}/\lbrace \pm \bone \rbrace$ is
generated by $T:\tau\to\tau+1$ and $S:\tau\to -1/\tau$,
and under the action of these two generators we have
$$ \Theta_{n,L}(\tau+1,z)~=~\exp\left\{2\pi i\,{{n^2}\over {4L}} \right\}
     \,\Theta_{n,L}(\tau,z)~,
\eqn\pThetaT
$$
and
$$ \Theta_{n,L}\left(-{1\over\tau},{z\over\tau}\right)~=~
    \sqrt{ {{-i\tau}\over{2n}} }
    \,
   \exp\left\lbrace {{\pi iLz^2}\over\tau}\right\rbrace \,
   \sum_{n'=0}^{2L-1}   \exp\left\lbrace {{-\pi inn'}\over L}\right\rbrace \,
      \Theta_{n',L}(\tau,z)~.
\eqn\pThetaS
$$
Here the square root indicates the branch with non-negative real part.
 From these results the modular transformation properties of the full
$SU(2)_K$ characters $\chi_\ell(\tau,z)$ can be obtained.

We are primarily interested in obtaining the characters $\calZ_{2m}^{2j}$
of the parafermion sectors $[\phi^j_m]$, for the string functions
$c^\ell_n$ are simply related to these characters via
$$ {\cal Z}^\ell_n(\tau)~=~\eta(\tau)\, c^\ell_n(\tau)~
\eqn\phichar
$$
where $\ell\equiv 2j$, $n\equiv 2m$, and where $\eta$ is the Dedekind
$\eta$-function:
$$    \eta(\tau) ~\equiv~ q^{1/24} \,\prod_{n=1}^\infty \,(1-q^n)~
      ~=~ \sum_{n=0}^\infty \,(-1)^n \,q^{3(n-1/6)^2/2}
\eqn\pDededef
$$
with $q\equiv \exp\lbrace 2\pi i \tau\rbrace$.
However, since the $\bZ_K$ parafermion theory is equivalent to the
$SU(2)_K/U(1)$ coset theory, the parafermion characters $\calZ_m^\ell$
can be obtained by expanding the full $SU(2)_K$ characters
$\chi_\ell(\tau,z)$ in a basis of $U(1)$ characters:\REF\kacpeterson{
   V.G.~Ka\v{c}, {\it Adv.~Math.} {\bf 35}
   (1980) 264; V.G.~Ka\v{c} and D.~Peterson, {\it Bull.~AMS}
    {\bf 3} (1980) 1057; {\it ibid.,} {\it Adv.~Math.} {\bf 53} (1984)
    125.}\refend\
$$   \chi_\ell(\tau,z) ~=~
   \sum_{n= -\ell}^{2K-\ell-1} \,
   \calZ_n^\ell(\tau)\,{{\Theta_{n,K}(\tau,z)}\over{\eta(\tau)}} ~=~
   \sum_{n= -\ell}^{2K-\ell-1} \, c^\ell_n (\tau) \,\Theta_{n,K}(\tau,z)~.
\eqn\pbasisexp
$$
Here $\Theta_{n,K}(\tau,z)/\eta(\tau)$ are the characters of a $U(1)$ boson
compactified at radius $2\sqrt{K}$.
Eq.~\pbasisexp, then, can be taken as a definition of the
string functions $c^\ell_n$.

Explicit expressions for the $c^\ell_n$ can be extracted from
\pSUtchars, \pJRtheta, and \pbasisexp.  The original formula
obtained by Ka\v{c} and Peterson\refmark{\kacpeterson}\ is
$$  c^\ell_n(\tau)~=~[\eta(\tau)]^{-3}\,
   {\sum_{x,y}}^\prime \,{\rm sign}(x)\,q^{x^2(K+2) -y^2 K}
\eqn\pkacpetoriginal
$$
where the prime on the summation indicates that $x$ and $y$ must
be chosen
so that three conditions are satisfied:
(1) $-|x|< y\leq |x|$; (2) either $x$ or $(\half -x)$ must equal
$(\ell+1)/[2(K+2)]$ modulo $1$; and (3) either $y$ or $(\half +y)$ must equal
$n/(2K)$ modulo $1$.
For many calculational purposes, however, a useful alternative expression
is\REF\Jacques{J.~Distler and Z.~Qiu, {\it Nucl.~Phys.~}
  {\bf B336} (1990) 533.}\refend\
$$  \eqalign{
    c^\ell_n(\tau) ~&=~ q^{h^\ell_n + [4(K+2)]^{-1}} \,\eta^{-3}\,
      \sum_{r,s=0}^\infty (-1)^{r+s}\,q^{r(r+1)/2\,+
      s(s+1)/2\,+rs(K+1)}\,\times\cr
    &~~\times\biggl\lbrace q^{r(j+m)\,+\, s(j-m)}~
      -~ q^{K+1-2j\,+\,r(K+1-j-m)\,+ \,s(K+1-j+m)}\biggr\rbrace\cr}
\eqn\pstringdef
$$
where $\ell-n\in 2{\bf Z}$ and where $h^\ell_n$ are the highest
weights given in \pfdims.  Note that the string functions
exhibit the symmetries
$$    c^\ell_n ~=~ c^\ell_{-n}~=~ c^{K-\ell}_{K-n} ~=~ c^\ell_{n+2K}~,
\eqn\pstringids
$$
as a consequence of which for any $K$ we are free to choose a ``basis''
of string functions $c^\ell_n$ where $0\leq \ell \leq K$ and $0 \leq n
\leq n_{\rm max}$, where $n_{\rm max}$ equals $\ell$ if $\ell\leq K/2$,
and $\ell-2$ otherwise.
Note also that the $K$-dependence of the string functions is suppressed in
this notation.  From \pstringdef, then, we see that the
string functions $c^\ell_n$ all take the general form
$q^{H^\ell_n}(1+...)$ where within the parentheses
all powers of $q$ are non-negative integers, and where
$$     H^\ell_n ~\equiv~ h^\ell_n ~+~ {1\over{4(K+2)}} ~-~
     {\textstyle{1\over 8}}
   ~=~ h^\ell_n ~-~ {K\over{8(K+2)}} ~.
\eqn\kstringqform
$$

For arbitrary fixed $K$, the set of
string functions forms an admissible representation of the modular group,
\ie, they close under modular transformations.
In fact, under $T$ they transform as eigenfunctions:
$$   c^\ell_n(\tau+1)~=~ \exp\bigl\{2\pi i H^\ell_n \bigr\}~c^\ell_n(\tau)~,
\eqn\pTtrans
$$
and under $S$ they mix among themselves:\refmark{\kacpeterson}\
$$  c^\ell_n(-1/\tau)~=~ {1\over{\sqrt{-i\tau}}}\,{1\over{\sqrt{K(K+2)}}}
    \,\sum_{\ell'=0}^K~\sum^K_{{n'= -K+1}
    \atop{\ell^\prime-n^\prime\in 2\bZ}}
    \,b(\ell,n,\ell',n') \,c^{\ell'}_{n'}(\tau)
\eqn\pStrans
$$
where the mixing coefficients $b(\ell,n,\ell',n')$ are
$$  b(\ell,n,\ell',n') ~=~ \exp\left\lbrace
    {{i\pi nn'}\over{K}} \right\rbrace \,\sin\left\lbrace
     {{\pi (\ell+1)(\ell'+1)}\over{K+2}} \right\rbrace ~.
\eqn\pbdef
$$
The first square root in \pStrans\ once again indicates the branch with
non-negative real part.
It is often convenient to define the linear combinations
$d^{\ell\pm}_n\equiv c^\ell_n \pm c^{K-\ell}_n$
when $\ell$ and $n$ are even:
if $K\in 4\bZ$, then these $d^\pm$-functions are also
eigenfunctions of $T$, and \pStrans\ implies that
that the $d^+$-functions close exclusively amongst themselves
under $S$.  The $d^-$-functions, on the other hand,
close exclusively with themselves and with the odd $(\ell,n)$ string
functions under $S$.  We will see that for $K>2$ all of our new identities
involve only the $d^\ell_n\equiv d^{\ell+}_n$ functions.

As expected, the string functions
$c^\ell_n$ reduce to the better-known Dedekind $\eta$-function \pDededef\
and the Jacobi $\vartheta_i$-functions
for the $K=1$ and $K=2$ special cases
respectively.  In particular,
for $K=1$ there is only one independent string function $c^0_0$,
and since the ``$\bZ_1$ parafermion'' theory $SU(2)_1/U(1)$ consists
of only the identity field $\phi^0_0=\bone$, we immediately find
$$     K=1:~~~~~ \calZ^0_0~=~\eta c^0_0~=~1 ~~~~\Longrightarrow~~~~
                c^0_0 ~=~ \eta^{-1}~.
\eqn\kKzerostringeta
$$
Similarly, for $K=2$, there are precisely three independent string functions
$c^0_0$, $c^1_1$, and $c^2_0$, and these can be expressed in terms of
the three non-vanishing Jacobi $\vartheta$-functions
$$  \eqalign{
      \thetatwo(\tau) ~&\equiv~ 2\,q^{1/8}\,\prod_{n=1}^\infty\,
                 (1+q^n)^2\,(1-q^n) ~\cr
      \thetathree(\tau) ~&\equiv~ \prod_{n=1}^\infty\,
                 (1+q^{n-1/2})^2\,(1-q^n) ~\cr
      \thetafour(\tau) ~&\equiv~ \prod_{n=1}^\infty\,
                 (1-q^{n-1/2})^2\,(1-q^n) ~\cr }
\eqn\pthetadefs
$$
via the relations
$$     K=2:~~~~~ \cases{
     2(c^1_1)^2     ~& $=~\thetatwo/\eta^3$  \cr
     (c^0_0+c^2_0)^2~& $=~\thetathree/\eta^3$  \cr
     (c^0_0-c^2_0)^2~& $=~\thetafour/\eta^3~.$  \cr}
\eqn\pstringtheta
$$
We thus see that for $K=1$ and $K=2$,
the string functions reproduce the boson and fermion characters respectively;
indeed, \pstringtheta\ reflects the fact that the $\bZ_2$
parafermion theory is equivalent to that of a free fermion in two dimensions.
For $K>2$, on the other hand, the string functions are the characters
of {\it para}\/fermions, and as such they
are not {\it a priori} related in these simple ways to the boson and
fermion characters.
However, we will see that our $[K,1]$ and $[K,2]$ series of
identities nevertheless provide such unexpected relations.
These new identities arise naturally in the partition functions of
fractional superstrings, and it is in such theories that they
find possible physical interpretations.

\section{Fractional Superstrings}

We now provide a brief introduction to the theory of
fractional superstrings.\refmark{\AT}\
Since our purpose here is to motivate
many of our new string-function identities and discuss
some of the physical contexts in which they arise, our treatment
will focus primarily on fractional-superstring partition functions.
A fuller treatment of these and other aspects of the fractional
superstring can be found in Refs.~[\AT] and
\REF\DT{K.R.~Dienes and S.-H.H.~Tye, {\it Model-Building
   for Fractional Superstrings}, McGill preprint
   McGill/91-29, Cornell preprint CLNS 91/1100, {\tt hepth@xxx/9112015}
   (November 1991).}
[\DT].

As indicated in the Introduction,
the basic idea behind the construction of the fractional
superstring is simple.\refmark{\AT}\  Let us first recall
the special case of the superstring.  The worldsheet structure of the
superstring theory is closely related to the $SU(2)_2$
WZW theory:\refmark{\Knizh}\
the worldsheet superpartner of the coordinate boson
$X^\mu$ is a Majorana fermion $\psi^\mu$ which can be described
by the $SU(2)_2/U(1)$ coset theory,
and the spacetime coordinate $X^\mu$ can be interpreted as
the remaining $U(1)$ boson but with its radius of compactification
relaxed to infinity.
(The spacetime index $\mu$ runs from $0$ to $D_c-1$.)
This boson-decompactification procedure destroys the $SU(2)_2$ symmetry
of the original WZW model, but its superconformal symmetry survives
and exists on the worldsheet.

The fractional superstring theory is related in
the same way to the $SU(2)_K$ WZW theory for $K\geq 2$.  The coset
theory $SU(2)_K/U(1)$ is the $\bZ_K$ parafermion theory,\refmark{\Zam,\Gep}\
and once again we obtain the spacetime coordinate field
$X^\mu$ by completely decompactifying the remaining WZW $U(1)$ boson.
Replacing the supercurrent for $K>2$ is a new chiral
current\REF\Bag{D.~Kastor, E.~Martinec, and Z.~Qiu, {\it
   Phys.~Lett.} {\bf 200B} (1988) 434;  J.~Bagger,
   D.~Nemeschansky, and S.~Yankielowicz, {\it Phys.~Rev.~Lett.}
   {\bf 60} (1988) 389; F.~Ravanini, {\it Mod.~Phys.~Lett.}
   {\bf A3} (1988) 397.  See also S.-W.~Chung, E.~Lyman, and
   S.-H.H.~Tye, {\it Fractional Supersymmetry and Minimal
   Coset Models in Conformal Field Theory}, Cornell preprint
   CLNS 91/1057, to appear in {\it Int.~J.~Mod.~Phys.}
   {\bf A7}  (1992).}\refmark{\Knizh,\Bag}\
whose conformal dimension is $(K+4)/(K+2)$;
these new currents have fractional spin, and transform the bosonic $X^\mu$
fields
to fractional-spin fields on the worldsheet.  It is natural to refer to this
remaining worldsheet symmetry as a fractional superconformal
symmetry,\REF\Jim{
    P.C.~Argyres, J.~Grochocinski, and S.-H.H.~Tye,
   {\it Nucl. Phys.} {\bf B367} (1991) 217;
   {\it ibid.}\/, {\it Construction of the $K=8$ Fractional
   Superconformal Algebras}, Cornell preprint CLNS 91/1126
   (January 1992).}\refend\
and to the strings based on these worldsheet
fractional supersymmetries as fractional superstrings.

We are interested in constructing the one-loop partition functions
$\calZ$ that such fractional superstring theories must have;
in particular, we focus here on the ``Type II'' fractional superstring
in which both the left-moving and right-moving worldsheet
theories exhibit a level-$K$ fractional supersymmetry.
We can therefore consider, for simplicity,
only the holomorphic components of $\calZ$;
these are the terms arising from the left-moving worldsheet degrees of
freedom.  As discussed in Sect.~2.1, for each $K$
such terms will be
products of the characters of free worldsheet bosons $X^\mu$ and
worldsheet $\bZ_K$ parafermions $(\phi^j_m)^\mu$:
each coordinate boson $X^\mu$
contributes to the partition function a factor of $1/\eta$
(the character of an infinite-radius boson), while its fractional
superpartner (the corresponding worldsheet parafermion)
contributes a factor of $\eta c^\ell_n$.  Thus,
the net holomorphic contribution to the fractional-superstring partition
function from each spacetime dimension is a factor of one string
function $c^\ell_n$.
Even though the fractional superstring theory is formulated in $D_c$
spacetime dimensions (\ie, even though the full worldsheet structure
of the fractional superstring is a tensor product of
$D_c$ copies of the individual boson/parafermion theories),
the large gauge symmetry of critical string theory is expected
to remove all time-like and longitudinal components, thereby leaving a
spectrum of physical states arising from the excitations of fields
corresponding
to only the $D_c-2$ transverse dimensions.\refmark{\AT,\DT}\
This is completely analogous to the case of ordinary $K=2$ superstring
theory, in which there are only $D_c-2=8$ ``effective''
dimensions giving rise to propagating fields (as is evident, for
example, in light-cone gauge).
The holomorphic components of $K$-fractional superstring partition
functions therefore consist of
$(D_c-2)$ factors of level-$K$ string functions, and take the general
form $c^{D_c-2}$.

Let us now consider the contribution to the partition
function from some of the low-lying states of the fractional superstring;
this will, as a byproduct, yield one method of determining the
critical dimension $D_c$.
Recall that in general an expansion $\sum_n a_n q^n$ of the
holomorphic factors within the partition function $\calZ$
indicates the net number $a_n$ of states with (left-moving) mass
$M^2=n$ contributing to that term.
As explained in Refs.~[\AT] and [\DT], the left-moving bosonic vacuum state
of the $D_c$-dimensional fractional superstring
corresponds to $(c^0_0)^{D_c-2}$.
Since \kstringqform\ indicates that this term takes the form
$$    (c^0_0)^{D_c-2}~\sim~ q^{H_{\rm tach}} \,(1+...)~~~
   ~~{\rm where}~~H_{\rm tach}~=~ -\,{(D_c-2)K\over8(K+2)}~,
\eqn\ptach
$$
we see that for $D_c>2$ the bosonic ground state is tachyonic.
This is analogous to the case of ordinary Type II superstrings,
in which the bosonic ground state is tachyonic with $M^2= H_{\rm tach}=-\half$.
Similarly, the left-moving vector state in
the fractional superstring theory contributes
to the first term in the expansion of
$$   (c^0_0)^{D_c-3}(c^2_0)~\sim~ q^{H_{\rm grav}}\,(1+...)~~~
   ~~{\rm where}~~H_{\rm grav}~=~{2\over K+2}\,-
     \,{{(D_c-2)K}\over{8(K+2)}}~.
\eqn\pvect
$$
Since this vector state is the left-moving component
of the graviton, the fractional superstring theory will therefore
be a theory of gravity (\ie, contain a massless graviton)
only if $H_{\rm grav}=0$, or
$$    D_c~-~2~=~ {16\over K}~.
\eqn\pcritD
$$
Thus, for $K=2,4,8$, and $16$ we have the integer critical spacetime
dimensions $D_c=10,6,4$, and $3$ respectively.

Physically, we are interested only in fractional superstring
theories which are tachyon-free.  This requires
that all tachyonic states be projected out of the physical spectrum, as occurs
in physically sensible superstring and heterotic string theories.  Hence,
when constructing  partition functions $\calZ_K$ for
the closed ``Type II'' $K$-fractional superstrings,
we seek modular-invariant combinations of terms of
the form $c^{D_c-2}=c^{16/K}$ in which the massless sector \pvect\
is present but the tachyonic sector \ptach\ is absent.

For the $K=2$ (superstring) theory, of course, there
exists such a unique tachyon-free modular-invariant
solution:\REF\GSO{
    F.~Gliozzi, J.~Scherk, and D.~Olive, {\it Nucl.~Phys.}
    {\bf B122} (1977) 253.  For a review see, \eg, M.B.~Green,
    J.H.~Schwarz, and E.~Witten, {\it Superstring Theory}, Vols.~1 and 2
    (Cambridge University Press, 1987).}
$$  \calZ_2(\tau)~=~\tautwo^{-4}\,|A_2|^2~.
\eqn\pZtwo
$$
Here $\tautwo$ is the imaginary part of $\tau$,
and $A_2$ is the modular-invariant combination
$$  A_2~=~8(c^0_0)^7(c^2_0) +56(c^0_0)^5(c^2_0)^3
 +56(c^0_0)^3(c^2_0)^5 +8(c^0_0)(c^2_0)^7 -8(c^1_1)^8.
\eqn\pAtwo
$$
This can be translated into a more familiar form by recalling
the equivalences \pstringtheta\ between the $K=2$ string functions and the
Jacobi $\vartheta$-functions;  using these results, we find that
$A_2$ can be rewritten as
$$  A_2~=~\half\,\eta^{-12}\,\left(
    \thetathree^4-\thetafour^4-\thetatwo^4\right)~=~\half\,\deltaminushalf\,J~
\eqn\pAJ
$$
where $\Delta\equiv \eta^{24}$ and $J$ is the vanishing Jacobi factor \kJdef.

We can construct solutions satisfying our requirements
for the $K>2$ cases as well.
For $K=4$, we find one tachyon-free modular-invariant partition
function\refmark{\AT}
$$  \calZ_4(\tau)~=~\tautwo^{-2}\,\left(|A_4|^2~+~12\,|B_4|^2\right)~
\eqn\pZfour
$$
where $A_4$ and $B_4$ are the combinations
$$  \eqalign{
  A_4~ = &~4(c^0_0+c^4_0)^3(c^2_0)-4(c^2_0)^4-4(c^2_2)^4
   +32(c^2_2)(c^4_2)^3~,\cr
  B_4 ~= &~8(c^0_0+c^4_0)(c^2_0)(c^4_2)^2
   +4(c^0_0+c^4_0)^2(c^2_2)(c^4_2)-4(c^2_0)^2(c^2_2)^2~.\cr}
\eqn\pABfour
$$
Unlike $A_2$, which was by itself modular-invariant,
$A_4$ and $B_4$ mix amongst themselves under modular transformations;
under $T$ we find that $A_4$ and $B_4$ transform
as eigenfunctions with eigenvalues $+1$ and $-1$ respectively,
whereas under $S$ we find
$$    S:~~~  \pmatrix{ A_4 \cr B_4 \cr} ~\longrightarrow~
      -\,{1\over{\tau^2}}\,\pmatrix{
      1/2 & 3 \cr   1/4 & -1/2 \cr} \,\pmatrix{A_4 \cr B_4 \cr}~.
\eqn\pABfourmix
$$

Similarly,
for the $K=8$ closed ``Type II'' fractional superstring,
we find the unique tachyon-free
modular-invariant partition function\refmark{\AT}\
$$  \calZ_8(\tau)~= ~\tautwo^{-1} \,\left(|A_8|^2~+~|B_8|^2~+~
            2\,|C_8|^2\right)~
\eqn\pZeight
$$
where we now have the three combinations
$$  \eqalign{
  A_8 ~= & ~2(c^0_0+c^8_0)(c^2_0+c^6_0)-2(c^4_0)^2
   -2(c^4_4)^2+8(c^6_4c^8_4)~,\cr
  B_8 ~= & ~4(c^0_0+c^8_0)(c^6_4)+4(c^2_0+c^6_0)(c^8_4)
   -4(c^4_0c^4_4)~,\cr
  C_8 ~= & ~4(c^2_2+c^6_2)(c^8_2+c^8_6)-4(c^4_2)^2~.\cr}
\eqn\pABCeight
$$
These combinations are eigenfunctions of $T$ with eigenvalues
$+1$, $-1$, and $-i$ respectively, and under $S$ they mix as follows:
$$ S:~~~  \pmatrix{ A_8 \cr B_8 \cr C_8 \cr }
      ~\longrightarrow~ {{i\over \tau}} \,
      \pmatrix{ 1/2 & 1/2 & 1 \cr
                1/2 & 1/2 & -1 \cr
                1/2 &-1/2 & 0 \cr}\,
 \pmatrix{ A_8 \cr B_8 \cr C_8 \cr }~.
\eqn\pABCeightmix
$$

For the $K=16$ fractional superstring there also exists a solution:
$$ \calZ_{16}~=~\tautwo^{-1/2}\,\left(|A_{16}|^2~+~4\,|C_{16}|^2\right)~,
\eqn\pZsixteen
$$
where our combinations are
$$ \eqalign{
    A_{16}~= & ~c^2_0+c^{14}_0-c^8_0-c^8_8+2c^{14}_8~,\cr
    C_{16}~= & ~c^2_4+c^{14}_4-c^8_4~.\cr}
\eqn\pABsixteen
$$
These combinations have eigenvalues $+1$ and $-i$ under $T$,
and under $S$ they mix as follows:
$$ S:~~~\pmatrix{A_{16}\cr C_{16}\cr}~\longrightarrow~{1\over{\sqrt{-i\tau}}}\,
    {1\over{2\sqrt{2}}} \,\pmatrix{ 2 & 4 \cr 1 & -2} \,
   \pmatrix{A_{16}\cr C_{16}\cr}~.
\eqn\pSsixteen
$$

It turns out that each of the partition functions $\calZ_K$ in this series
has a remarkable property:
viewed as a function of $q\equiv e^{2\pi i \tau}$, each vanishes identically.
We shall prove this assertion in Sect.~4.
In the $K=2$ (superstring) case, we see from \pZtwo\ and \pAJ\
that this vanishing is equivalent to the Jacobi identity
$$  J~\equiv~\thetathree^4~-~\thetatwo^4 ~-~\thetafour^4~=~0~,
\eqn\pJacobi
$$
and indeed it is well-known that the particle spectrum of this
Type II superstring exhibits a spacetime supersymmetry.
Thus, the famous Jacobi identity \pJacobi\ is the reflection
(at the partition-function level) of this underlying spacetime supersymmetry.
In analogous fashion, we interpret the vanishing of each $\calZ_K$
for $K>2$ as a sign of spacetime supersymmetry
in the {\it fractional}\/-superstring spectrum of states:
the contribution to $\calZ_K$ from each bosonic state at every mass level
in the theory is cancelled by the (equal but opposite) contribution
from a corresponding fermionic state.
Since these $K>2$ partition functions $\calZ_K$ have been written as
the sums of squares,
it follows that the separate string-function
combinations $A_K$, $B_K$, and $C_K$ must each
independently also vanish as functions of $q$:
$$   A_4~=~B_4~=~A_8~=~B_8~=~C_8~=~A_{16}~=~C_{16}~=~0~.
\eqn\pABCvanishings
$$
These string-function identities \pABCvanishings\ are therefore the
$K>2$ analogues of the ($K=2$) Jacobi identity,
and we shall prove this series of new Jacobi-like identities in Sect.~4.

We emphasize
that in spite of \pABCvanishings, one-loop modular invariance continues
to require that the individual terms $|A_K|^2$, $|B_K|^2$, and
$|C_K|^2$ appear together in our partition functions $\calZ_K$
in the combinations given in \pZfour, \pZeight, and \pZsixteen.
As in the superstring, one-loop modular invariance not only
guarantees multi-loop modular invariance, but is also
required for the internal consistency of the theory.
In fact, since \pABCvanishings\ is the reflection of spacetime
supersymmetry in the fractional-superstring spectrum,
such cancellations must be distinguished from those due to ``internal''
(GSO-like) projections which act between states with
the {\it same}\/ statistics and which thereby
actually {\it remove} physical states from the theory.

Given that these partition functions $\calZ_K$ vanish,
we now turn to an examination of the individual terms within
each $\calZ_K$.  In particular, since this vanishing is a
reflection of spacetime supersymmetry, we focus on determining
the spacetime statistics of the particles contributing to each
of the terms in $\calZ_K$.

Let us first recall the well-understood $K=2$ case.
In the $K=2$ superstring partition function (where we are restricting
ourselves to the left-moving holomorphic sector only),
we have seen that the expression $A_2$ is proportional to
$\deltaminushalf J$, and it is well-known that within this
factor $\deltaminushalf J$ the terms
$\deltaminushalf(\thetathree^4-\thetafour^4)$ represent the
contributions from spacetime bosonic states (\ie, from the
worldsheet Neveu-Schwarz sector).  Likewise, the remaining
term $\deltaminushalf\thetatwo^4$ within $\deltaminushalf J$
represents the contributions from spacetime fermionic states (\ie, from
the worldsheet Ramond sector).
Using the relations \pstringtheta\ between the Jacobi
$\vartheta$-functions and the string functions, we see
that the fermionic Ramond sector contributes only to the term $(c^1_1)^8$
within $A_2$, while the remaining terms within $A_2$ receive contributions from
only the bosonic Neveu-Schwarz sector.
Having thus distinguished these contributions in terms of the $K=2$
string functions, we can now easily discern which of the
worldsheet $\bZ_2$ ``parafermion'' primary fields $\phi^j_m$
are responsible for bosonic or fermionic statistics:
since the character of each field $\phi^j_m$ is $\eta c^{2j}_{2m}$,
we see that the parafermion fields giving rise to spacetime fermionic
states are $\phi^{1/2}_{\pm1/2}$, while
those giving rise to spacetime bosonic states are
$\phi^j_m$ with quantum number $m=0$.
Thus, in light-cone gauge, spacetime
bosons will have vertex operators proportional to
$$  B~\sim~\prod_{i=1}^{8}\left(\phi^{j_i}_0\right)~
\eqn\ptwobosons
$$
where $j_i=0$ or $1$, and
spacetime fermions will have vertex operators proportional to
$$  F~\sim~\prod_{i=1}^{8}\left(\phi^{1/2}_{m_i}\right)~
\eqn\ptwoferms
$$
where $m_i = \pm 1/2$.
Note that we are suppressing the
contributions to the vertex operators from the worldsheet bosons $X^\mu$.
Since the worldsheet bosons give rise only to
states with spacetime bosonic statistics, this suppression
does not affect the identification of the statistics of
vertex operators.
Similarly, any of the above parafermion primary fields $\phi^j_m$ may
be replaced by one of its descendant fields without altering the statistics.

It is straightforward to demonstrate that the above vertex operator
assignments satisfy a number of self-consistency checks.
First, we can recall the correspondence between the $\bZ_2$ parafermion
fields and the free fermion fields of the Ising model
$$ \eqalign{
   \phi^0_0~=~\bone~,\ \ \ \ \quad\quad\quad\phi^1_0~&=~\psi~,\cr
   \phi^{1/2}_{1/2}~=~\sigma~,\quad\quad\quad
   \phi^{1/2}_{-1/2}~&=~\sigma^{\dagger}~;\cr}
\eqn\pKtwotoferm
$$
here $\psi$ is the Majorana fermion field and
$\sigma$ and $\sigma^{\dagger}$ are the spin fields.
These worldsheet spin fields and their descendants create the
(Ramond) spacetime fermionic states from the vacuum, and the Majorana and
identity fields similarly create the (Neveu-Schwarz) spacetime bosonic
states.  This is therefore consistent with the above vertex-operator
assignments.  More compellingly, we can check that
the $\bZ_2$ parafermion algebra itself reproduces
the correct spin-statistics selection rules.
Recall that according to the fusion rules \pfusion,
the $m$-quantum number of the parafermion field is additive.
 From this and the field identifications \phids, it follows that
our vertex operators \ptwobosons\ and \ptwoferms\ satisfy the following
selection rules under fusion:
$$ \eqalign{
    F\times F~&\to~B~\cr
    F\times B~&\to~F~\cr
    B\times B~&\to~B~.\cr}
\eqn\pstats
$$
This is, of course, in accordance with the required spin-statistics
connection.

These considerations can easily be generalized to the $K>2$ cases:
here the analogues of \ptwobosons\ and \ptwoferms\ are
$$    B~\sim~\prod_{i=1}^{D_c-2}\left(\phi^{j_i}_0\right)~
    \quad\quad{\rm and}\quad\quad
  F~\sim~\prod_{i=1}^{D_c-2}\left(\phi^{j_i}_{\pm K/4}\right)~,
\eqn\pKfermbos
$$
and from the equivalences \phids\ and the fusion rules \pfusion\
we see that the selection rules \pstats\ are again satisfied.

This identification of the spacetime spin and statistics of
fractional-super\-string states is compatible with the $A_K$
parts of the partition functions $\calZ_K$ given above, for
each term within each $A_K$ can readily be identified as the
contribution from either a spacetime bosonic state or a
spacetime fermionic state.  It is therefore straightforward to
decompose each $A_K$ into two pieces (just as was done for
the $K=2$ Jacobi factor $J$), writing
$$    A_K ~=~ A_K^b ~-~ A_K^f ~
\eqn\kAKsplit
$$
where $A_K^b$ (respectively $A_K^f$) contains the terms
in which all string functions are of the form $c^\ell_0$
(respectively $c^\ell_{K/2}$):
$$  \eqalign{
   K=2:~~~~A_2^b ~&\equiv
         ~8(c^0_0)^7(c^2_0) +56(c^0_0)^5(c^2_0)^3
         +56(c^0_0)^3(c^2_0)^5 +8(c^0_0)(c^2_0)^7  \cr
   A_2^f ~&\equiv ~8(c^1_1)^8~  \cr
   K=4:~~~~A_4^b ~&\equiv~ 4(c^0_0+c^4_0)^3(c^2_0)-4(c^2_0)^4  \cr
   A_4^f ~&\equiv ~4(c^2_2)^4-32(c^4_2)^3(c^2_2)~  \cr
   K=8:~~~~
   A_8^b ~&\equiv~ 2(c^0_0+c^8_0)(c^2_0+c^6_0)-2(c^4_0)^2  \cr
   A_8^f ~&\equiv ~2(c^4_4)^2-8(c^6_4)(c^8_4)~  \cr
   K=16:~~~~
   A_{16}^b ~&\equiv~ c^2_0+c^{14}_0-c^8_0  \cr
   A_{16}^f ~&\equiv ~c^8_8-2c^{14}_8~.  \cr}
\eqn\kstringfctssplit
$$
This separation of terms is also consistent with our previous observations.
First, the term we identified as containing a massless vector particle
in \pvect\ is of the form $\prod_i(c^{\ell_i}_{0})$, in
agreement with the form of the boson vertex operators given in
\pKfermbos;  note that it appears with a positive sign in $A_K^b$.
Next, we observe that according to the above
identification, massless fermions
must correspond to the first term in the $q$-expansion of
$$  \left(c^{K/2}_{K/2}\right)^{16/K}~\sim~q^0\,(1+...)~;
\eqn\pfermexp
$$
fortunately this term is present within each of our expressions $A_K^f$,
and indeed it can be shown\refmark{\AT}\ that the physical-state
conditions on the massless state in \pfermexp\ yield the
massless Dirac equation.  The Fermi statistics carried by
these spacetime spinorial states is reflected in the
negative sign with which this term appears in each $A_K$ (or the positive
sign within each $A_K^f$), and in fact
the coefficient for this term in each case gives precisely
the counting of physical states necessary for a Majorana and/or Weyl
fermion in the critical spacetime
dimension $D_c$ of the fractional superstring.
Furthermore, note that this left-moving spin-1/2 massless fermionic state
in \pfermexp\ can be tensored with the right-moving spin-1 massless
bosonic state in \pvect\ (and {\it vice versa}\/) to form a massless
spin-3/2 gravitino.  Consistency would then require that the closed
fractional-superstring spectrum exhibit an $N=2$ spacetime supersymmetry.
The vanishing of each $A_K$, $B_K$, and $C_K$ (or the equality of their
respective bosonic and fermionic parts) is indeed consistent with this
conclusion.

One new feature for $K>2$ is the presence of additional terms within
$A_K^b$ and $A_K^f$ which appear with the ``wrong'' signs (\ie,
negative signs within the individual pieces $A_K^b$ and $A_K^f$).
Although these flipped signs might initially seem to contradict our
identification of the statistics of states, it turns out that
the expressions $A_K^f$ as defined in \kstringfctssplit\ satisfy
the following identities:
$$\eqalign{
K=2:\quad\quad\quad  A^f_2~&
   =~8\left(\prod_{n=1}^\infty{1+q^n\over1-q^n}\right)^8~\cr
 K=4:\quad\quad\quad A^f_4~&
   =~4\left(\prod_{n=1}^\infty{1+q^n\over1-q^n}\right)^4~\cr
 K=8:\quad\quad\quad A^f_8~&
   =~2\left(\prod_{n=1}^\infty{1+q^n\over1-q^n}\right)^2~\cr
 K=16:\quad\quad\quad A^f_{16}~&
   =~\phantom{2}\left(\prod_{n=1}^\infty
     {1+q^n\over1-q^n}\right)~.\cr}
\eqn\ptherms
$$
These new identities will be proven in Sect.~4.
Note that since $A_K=A_K^b-A_K^f=0$, the bosonic
terms $A_K^b$ also have the above $q$-expansions.
Since \ptherms\ implies that the coefficients in the
$q$-expansions of each $A_K^b$ and $A_K^f$ are all positive,
we are led to view the minus signs in
$A^b_K$ and $A_K^f$ as ``internal'' projections (or cancellations)
of degrees of freedom in the fractional superstrings.
Furthermore, the pattern inherent in this series of identities
(matching smoothly as it does onto the well-understood $K=2$ case)
also suggests that our statistics identification is indeed correct.

 From the definitions of $\eta$ and $\vartheta_2$,
Eqs.~\pDededef\ and \pthetadefs\ respectively, it follows that
$$  \left(\prod_{n=1}^\infty {{1+q^n}\over{1-q^n}}\right)^2
    ~=~{\thetatwo\over{2\,\eta^3}}~=~ (c^1_1)^2~
\eqn\pthetatwopartition
$$
where $c^1_1$ is a level $K=2$ string function.
Thus the identities \ptherms\ can be viewed as relating the $K>2$
parafermion characters (string functions) to $K=2$ fermion characters
(Jacobi $\vartheta$-functions), and indeed these identities
are just some of the identities in our $[K,2]$ series.
The full set is collected in Sect.~3.

One immediate consequence of \ptherms\ is that
the level-by-level counting of physical degrees of freedom
in each of the fermionic $A_K^f$ sectors in the fractional
superstring is identical to that of the
Ramond sector of the usual superstring,
except with spacetime dimension $D_c=2+16/K$ instead of $10$.
At first glance, this observation might seem to imply that
the $A_K$ spectrum of the fractional superstring can itself
be equivalently described by the worldsheet
Majorana-fermion and free-boson theories of the ordinary ($K=2$) superstring,
without any need for worldsheet parafermions.
Indeed, the spectrum of the spacetime {\it fermionic}\/
sectors $A_K^f$ of the fractional superstring can be generated
by $D_c-2=16/K$ pairs of free worldsheet bosons $X^\mu$ and worldsheet
Ramond fermions $\psi^\mu$, as \ptherms\ and \pthetatwopartition\ jointly
suggest, and this is of course simply the operator
content of the Ramond sector in ordinary superstring theory.
However, consider the counting of the spacetime {\it bosonic}\/ states
(Neveu-Schwarz sector) for the superstring
$$  K=2:\quad\quad A^b_2~\equiv~
    \half \left[(c^0_0+c^2_0)^8-(c^0_0-c^2_0)^8\right]
    ~=~8\left({{\thetathree^4-\thetafour^4}
    \over 16\,\eta^{12}}\right)~.
\eqn\pKtwoNS
$$
Here $\sqrt{\thetathree/\eta}$ and $\sqrt{\thetafour/\eta}$ are the
characters of worldsheet Majorana fermions obeying Neveu-Schwarz boundary
conditions, and it is for this reason that we have written
$\thetathree^4-\thetafour^4$ rather than $\thetatwo^4$
in \pKtwoNS.  From \ptherms, then,
we can write the $K>2$ generalizations of \pKtwoNS\ as
$$\eqalign{
K=4:\quad\quad A^b_4~=~&
    ~4\left({{\thetathree^4-\thetafour^4}
    \over 16\,\eta^{12}}\right)^{1/2}\cr
K=8:\quad\quad A^b_8~=~&
    ~2\left({{\thetathree^4-\thetafour^4}
    \over 16\,\eta^{12}}\right)^{1/4}\cr
K=16:\quad\quad A^b_{16}~=~&
    ~\phantom{2}\left({{\thetathree^4-\thetafour^4}
    \over 16\,\eta^{12}}\right)^{1/8}~.\cr}
\eqn\pNStherm
$$
The right sides of \pNStherm\ cannot be interpreted as the characters
of any tensor products of bosons and Majorana fermions because of
the presence of the fractional exponents for $K>2$.
Therefore, an attempted description
of the $A_K$ part of the fractional-superstring spectrum
in terms of worldsheet bosons and fermions fails,
and indeed it is only the introduction of {\it parafermions}
on the worldsheet which allows us to generate the bosonic-sector
partition functions given in \pNStherm.

Thus far we have focused exclusively on the $A_K$ sectors
of the fractional superstrings, and we have seen that
obtaining the desired fractional exponents in \pNStherm\
(\ie, reducing the critical dimension from $D_c=10$ to $D_c=2+16/K$)
is achieved through the introduction of parafermions on the
string worldsheet.
There is, however, a price that must be paid.
Let us now rewrite the entire expression $A_K=A_K^b-A_K^f$ in terms of
Jacobi $\vartheta$-functions:  using \pNStherm\ for $A_K^b$, and
using \ptherms\ and \pthetatwopartition\ jointly for $A_K^f$, we find
$$    A_K~\propto ~\Delta^{-1/K} \,
     \biggl\lbrace
       \left(\thetathree^4-\thetafour^4\right)^{2/K} ~-~ \thetatwo^{8/K}
     \biggr\rbrace~
\eqn\kAKreconstructed
$$
where $\Delta\equiv \eta^{24}$.
Thus, we see that only for $K=2$ is $A_K$ by itself modular-invariant;
for $K>2$ we find that $A_K$ does not close into itself under $S$,
but rather requires the introduction of additional
sectors (such as those giving rise to $B_K$ and $C_K$)
in order to achieve modular-invariant partition functions.
This is the underlying reason why these additional expressions appeared
naturally in our partition functions $\calZ_K$.
Since these new sectors contain only massive states (\ie, states
with masses at the Planck scale), we consider
their introduction a small price to pay for the ability to decrease
the critical spacetime dimension in string theory.
Furthermore, we have seen that these additional expressions $B_K$ and $C_K$
also vanish as functions of $q$, and therefore the introduction
of these additional sectors preserves the spacetime supersymmetry at
all mass levels of the theory.

These additional sectors, however, seem to contain
much of the spacetime
physics which is intrinsically new to fractional superstrings.
To see this, let us consider the spacetime statistics of
the states appearing in these sectors.
The fields corresponding to terms in the $B_K$ sector are
themselves products of parafermion fields, half of which
have quantum number $m=0$ and half of which
have quantum number $m=K/4$.  Therefore, according to our previous discussion,
their vertex operators will have the following form
in light-cone gauge:
$$ Q_B~\sim~\prod_{i=1}^{D_c/2-1}\left(\phi^{j_i}_0\,
   \phi^{j_i'}_{\pm K/4}\right)~.
\eqn\pBKquarters
$$
According to \pKfermbos, these states
are therefore neither fermions nor bosons in $D_c$ spacetime dimensions.
A similar situation exists for the $C_K$ sectors.
These sectors consist of
terms of the form $\prod_i(c^{j_i}_{K/4})$ [where we
recall the identities \pstringids], and in light-cone gauge
these naturally correspond to vertex operators of the form
$$ Q_C~\sim~\prod_{i=1}^{D_c-2}\left(\phi^{j_i}_{\pm K/8}\right)~.
\eqn\pCKquarters
$$
Once again, such states cannot be interpreted
as bosons or fermions in $D_c$ spacetime dimensions.

In order to gain some insight into
the properties of these states, we can calculate the fusion
rules that these additional vertex operators $Q_B$ and $Q_C$ satisfy.
It turns out that these rules depend on the level $K$ of the fractional
superstring, since the number of parafermion fields within each
vertex operator depends on the critical dimension $D_c$ (and hence on $K$).
For the $K=4$ superstring, we have only the $A_4$ and $B_4$ sectors:
the $A_4$ sector can be decomposed into bosonic and fermionic pieces satisfying
the algebra \pstats, as we have seen, and the $B_4$ sector introduces
the additional vertex-operator fusion rules:
$$       \eqalign{
    Q_B\times B~&\to~Q_B~\cr
    Q_B\times F~&\to~Q_B~\cr
    Q_B\times Q_B~&\to~B~{\rm or}~F~ {\rm or}~Q_B~.\cr }
\eqn\pquarterstats
$$
These selection rules are therefore suggestive of ``spin-quarter''
statistics for $B_4$-sector particles, since the fusing of two identical
$B_4$-sector particles can result in an $A_4$-sector fermion.
Similarly, for the $K=8$ superstring, we have three sectors:
$A_8$, $B_8$, and $C_8$.  While the $A_8$ sector again leads to the fusion
rules \pstats, the $B_8$ sector now introduces the (slightly modified)
fusion rules:
$$       \eqalign{
    Q_B\times B~&\to~Q_B~\cr
    Q_B\times F~&\to~Q_B~\cr
    Q_B\times Q_B~&\to~B~{\rm or}~F~ \cr }
\eqn\kQBeight
$$
and the $C_8$ sector introduces the additional rules:
$$ \eqalign{
    Q_C \times B ~&\to~Q_C \cr
    Q_C \times F ~&\to~Q_C \cr
    Q_C\times Q_B~&\to~Q_C~\cr
    Q_C\times Q_C~&\to~B~{\rm or}~F~{\rm or}~Q_B~.\cr}
\eqn\peighthstats
$$
Once again, therefore, the $B_8$-sector fusion rules suggest
``spin-quarter'' statistics, and the $C_8$-sector rules seem to
indicate ``spin-eighth'' statistics.
For the $K=16$ string, on the other hand, there are only an $A_{16}$
and a $C_{16}$ sector.  In this case the vertex operators contain
only one parafermion field, and while the $A_{16}$ sector again
leads to \pstats, the $C_{16}$ sector now yields the fusion rules:
$$       \eqalign{
    Q_C\times B~&\to~Q_C~\cr
    Q_C\times F~&\to~Q_C~ \cr
    Q_C\times Q_C~&\to~B~{\rm or}~F~. \cr}
\eqn\kQBeight
$$
Thus in this instance it is the $C_{16}$ sector which
seems to suggest ``spin-quarter'' statistics.

Even though this discussion
has focused on only the left-moving sectors of the fractional
superstrings, these considerations indicate that it is likely
that the $B_K$ and $C_K$ sectors break the usual spin-statistics
connection in the critical dimension.  (This is not necessarily true
of the $K=16$ string, since fractional statistics are
allowed in three spacetime dimensions.)
This means that, for the fractional superstrings to
be consistent, either Lorentz invariance, quantum mechanics, or locality
must be broken in some way in the critical dimension.

An important point following from all of these
fusion rules is that the tree-level scattering
of particles in the $A_K$ sector can involve only other fields
in the $A_K$ sector as intermediate states.  Since only the $A_K$
sectors contain the massless states, we see that the $B_K$ and $C_K$
sectors make no contribution to the semi-classical
low-energy physics of the fractional superstrings.  Therefore,
since Lorentz invariance and the spin-statistics
connection appear to hold
in the $A_K$ sector, the semi-classical low-energy
physics predicted by the fractional superstrings
will indeed be the familiar Yang-Mills and gravity theories
(plus corrections proportional to powers of the
string tension, as in the usual superstring).
At the string loop level, however, fields in the $B_K$ or $C_K$
sectors can contribute to the scattering amplitudes of the massless
particles in the $A_K$ sector.  Thus, it is only the quantum
effects of the fractional superstring which render the massless (\ie,
potentially observable) physics inconsistent with the traditional
spin-statistics connection.  Since the lowest-lying states in
the $B_K$ and $C_K$ sectors have masses at the Planck scale, we
expect the quantum corrections to the massless-sector gravity
and Yang-Mills theories to be suppressed by factors of the
Planck mass relative to the low-energy scale (at least for
sufficiently weak string coupling).

These violations of the spin-statistics connection, though
potentially weak, may nevertheless be important qualitative signals of
stringy behavior.  Indeed, there are many possible mechanisms
through which these $B_K$ and $C_K$ sectors might lead to violations
of the usual spin-statistics relation.  First,
as mentioned above, the spin-statistics connection can be
invalidated by sacrificing locality.  Since strings are extended
objects, it is possible that the massive states in the $B_K$
and $C_K$ sectors correspond to extended states that cannot
be interpreted as elementary particles.  Indeed, the possibility
of exotic statistics due to the extended nature of strings
has been discussed in
  \REF\Bal{C.~Aneziris, A.P.~Balachandran, L.~Kauffman, and
  A.M.~Srivastava, {\it Int. J. Mod. Phys.} {\bf A6} (1991) 2519;
  J.A.~Harvey and J.~Liu, {\it Phys. Lett.} {\bf B240} (1990) 369.}
Ref.~[\Bal].
This proposal relies on the non-trivial nature of the motion
group in string-configuration space in four spacetime dimensions.
Another possibility is
that the states in the $B_K$ and $C_K$ sectors are soliton-like objects,
extended objects formed from the $A_K$-sector fields.
The slow fall-off of massless scalar or vector fields could
allow the violation of the usual
statistics selection rules, similar to what occurs in monopole-fermion
systems.\REF\monopole{
    A.S.~Goldhaber, {\it Phys.~Rev.~Lett.} {\bf 36} (1976)  1122.}\refend\

On the other hand, the spin-statistics
theorem can be avoided at the expense of Lorentz invariance.
Lorentz invariance is a symmetry known to hold only at
distances much larger than the Planck scale.  As we have
seen, these fractional superstring theories lead
to a breaking of Lorentz invariance only at the Planck scale and
only at the quantum loop level.  Such a breaking may actually exist,
and simply not be observable experimentally at the present time.

Alternatively, even if the massive-sector breaking of Lorentz
invariance in the critical dimension $D_c$ is strong even at low
energies, a lower-dimensional Lorentz invariance might still survive.
In this scenario, the spacetime symmetry group of the $B_K$ and $C_K$ sectors
would have a $D$-dimensional Lorentz subgroup with $D<D_c$,
providing a mechanism for spontaneously compactifying from
the critical dimension $D_c$ down to $D$ dimensions.
Let us see explicitly how this might occur.\refmark{\DT}\
Recall the form of the $B_K$-sector vertex operators given in
\pBKquarters.  While we have seen that these states cannot be interpreted
as either bosons or fermions in $D_c$ spacetime dimensions, they may well
have bosonic or fermionic interpretations in spacetime dimensions $D<D_c$.
For example, a bosonic interpretation for $Q_B$ is possible if the
parafermion fields with quantum number $m=0$ within \pBKquarters\ are viewed
as the (fractional) superpartners of the $D<D_c$ worldsheet coordinate bosons,
with the remaining $m\not= 0$ parafermion fields viewed as part of an
``internal'' worldsheet theory resulting from spacetime compactification.
In fact, it is
shown in Ref.~[\DT] that a theory fully consistent with Lorentz invariance and
the spin-statistics connection is possible for the $K=4$ and $K=8$
cases if the spacetime dimensions are compactified respectively to $D=4$
and $D=3$.
Furthermore, such a compactification scheme in the $K=4$ case
may also permit the construction of four-dimensional
fractional-superstring models containing chiral
spacetime fermions.\refmark{\DT}\

It is therefore evident that there is much potentially new
physics to be discovered
within fractional superstring theory, whether a possible
Planck-scale breakdown of Lorentz invariance, the appearance
of exotic statistics due to the extended (non-local) nature of strings,
or a ``self-compactification'' required for internal
consistency.  While the interpretations of many of these
effects have yet to be resolved,
the important point is that fractional superstrings
offer a unique and concrete framework in which to explore these issues.

\endpage

\chapter{LIST OF NEW STRING-FUNCTION IDENTITIES}

In this section we gather together all of the new string-function
identities to be proven in Sect.~4.  As discussed in the Introduction,
we refer to an identity as a $[K_1,K_2]$ identity
if it relates string functions $c^\ell_n$ of level $K_1$
to string functions of level $K_2$.  Recall that the
level $K=1$ string function is equivalent to the Dedekind $\eta$-function,
and that the level $K=2$ string functions are equivalent to the Jacobi
$\vartheta_i$-functions [these relations are given in \kKzerostringeta\
and \pstringtheta].
If we define the level $K=0$ function $c^0_0 \equiv 0$,
then our new identities come in three distinct series:
these are the $[K,0]$,
$[K,1]$, and $[K,2]$ series, for $K=2,4,8$, and $16$.
These identities are listed below.

\section{First Series: The $[K,0]$ Identities}

This series of identities generalizes the famous
Jacobi ``supersymmetry'' identity on fermion characters:
$$  J~\equiv~\thetathree^4~-~\thetatwo^4~-~\thetathree^4 ~=~0~.
\eqn\theJacobi
$$
In terms of $K=2$ string functions, \theJacobi\ is equivalent to
$$  \eqalign{
   A_2~&\equiv~8\,(c^0_0)^7(c^2_0)~+~56\,(c^0_0)^5(c^2_0)^3~+~
   56\,(c^0_0)^3(c^2_0)^5~+~8\,(c^0_0)(c^2_0)^7~-~8\,(c^1_1)^8\cr
      &=~ \half \,\deltaminushalf\,J~=~0\cr}
\eqn\AKequaltwo
$$
where $\Delta\equiv \eta^{24}$.
The Jacobi identity \theJacobi\ can therefore be regarded as
the $[2,0]$ special case,
and the analogous $[K,0]$ Jacobi-like ``supersymmetry'' identities for the
parafermion characters at higher levels $K$ are as follows.
For $K=4$, we define
$$ \eqalign{
  A_4~&\equiv~4\,(c^0_0~+~c^4_0)^3\,(c^2_0)~-~4\,(c^2_0)^4
   ~-~4\,(c^2_2)^4~+~32\,(c^2_2)\,(c^4_2)^3~,\cr
  B_4~&\equiv~8\,(c^0_0~+~c^4_0)\,(c^2_0)(c^4_2)^2
   ~+~4\,(c^0_0~+~c^4_0)^2\,(c^2_2)(c^4_2)
   ~-\,4\,(c^2_0)^2\,(c^2_2)^2~, \cr}
\eqn\ABKequalfour
$$
and for $K=8$, we define
$$\eqalign{
  A_8~&\equiv~2\,(c^0_0~+~c^8_0)\,(c^2_0~+~c^6_0)~-~2\,(c^4_0)^2
   ~-~2\,(c^4_4)^2~+~8\,(c^6_4c^8_4)~,\cr
  B_8~&\equiv~4\,(c^0_0~+~c^8_0)(c^6_4) ~+~ 4\,(c^2_0~+~c^6_0)\,(c^8_4)
   ~-~4\,(c^4_0c^4_4)~,\cr
  C_8~&\equiv~4\,(c^2_2~+~c^6_2)\,(c^8_2~+~c^8_6)~-~4\,(c^4_2)^2~.\cr }
\eqn\ABCKequaleight
$$
Similarly, for $K=16$, we define
$$  \eqalign{
    A_{16}~&\equiv~c^2_0~+~c^{14}_0~-~c^8_0~-~c^8_8~+~2\,c^{14}_8~,\cr
    C_{16}~&\equiv~c^2_4~+~c^{14}_4~-~c^8_4~.\cr}
\eqn\ABKequalsixteen
$$
Then each of these expressions also vanishes as a function of $q$:
$$  A_2~=~A_4~=~B_4~=~A_8~=~B_8~=~C_8~=~A_{16}~=~C_{16}~=~0~.
\eqn\paraJacobi
$$
These new ``Jacobi-like'' identities therefore form the $[K,0]$ series.
Note that while the Jacobi identity is the {\it unique}\/ such
identity for $K=2$, for each level $K>2$ there are in fact {\it several}\/
independent Jacobi-like $[K,0]$ identities.

\section{Second Series: The $[K,2]$ Identities}

This series of identities relates the $\bZ_K$ parafermion characters
to the ordinary fermion characters, and are the higher-$K$ analogues
of the $K=2$ ``identities''
$\deltaminushalf{\vartheta_i}^4 =\deltaminushalf{\vartheta_i}^4$
for $i=2,3,4$.

For $K=4$, we define the following quantities:
$$  \eqalign{
    A_4^b~&\equiv~ 4\,(d^0_0)^3 c^2_0 ~-~ 4\,(c^2_0)^4 \cr
    A_4^f~&\equiv~ 4\,(c^2_2)^4 ~-~ 32\,(c^4_2)^3 c^2_2 \cr
     C_4 ~&\equiv~ 4\,c^2_0 (c^2_2)^3 ~-~12 \,d^0_0 (c^4_2)^2 c^2_2
         ~-~ 8\,c^2_0 (c^4_2)^3 ~=~ 4\,q^{1/4}\,(1+...) \cr
     D_4 ~&\equiv~ (d^0_0)^3 c^2_2 ~+~ 6\,(d^0_0)^2 c^2_0 c^4_2
              ~-~ 4\,(c^2_0)^3 c^2_2 ~=~ q^{-1/4} \,(1+...) ~,\cr}
\eqn\pbfsplitfourdefs
$$
where $d^\ell_n\equiv c^\ell_n+c^{K-\ell}_n$.
Note that $A_4^b-A_4^f=A_4$.
Then our $[4,2]$ identities, which contain the $[4,0]$ Jacobi-like
identities $A_4=B_4=0$ as a subset, are as follows:
$$  \eqalign{
     A_4^b ~&=~ \thetatwo^2/\eta^6 \cr
     A_4^f ~&=~ \thetatwo^2/\eta^6 \cr
              B_4 ~&=~ 0 \cr
       C_4~&=~ \half\, (\thetathree^2-\thetafour^2)/\eta^6 \cr
       D_4~&=~ \half\, (\thetathree^2+\thetafour^2)/\eta^6 ~.\cr }
\eqn\pbfsplitfour
$$
As a consequence of \pbfsplitfour, we have
$$   (A_4^b+A_4^f)^2 ~-~ 16\,C_4\,D_4 ~=~ 4\,\Delta^{-1/2}\,J~=~0 ~.
\eqn\pbfcorollary
$$
Similarly, for $K=8$, we define
$$  \eqalign{
     A_8^b ~&\equiv ~2\,d^0_0 d^2_0 ~-~ \half \,(d^4_0)^2 \cr
     A_8^f ~&\equiv ~\half\,(d^4_4)^2 ~-~ 2\,d^6_4 d^8_4 \cr
     E_8 ~&\equiv~ \half\,d^4_2 d^4_4 ~-~ d^2_2 d^8_4 ~-~ d^6_4 d^8_2
           ~=~2\, q^{3/8}\,(1+...) \cr
     F_8 ~&\equiv~ \half\,d^0_0 d^2_2 ~+~ \half \,d^2_0 d^8_2 ~-~ d^4_0
         d^4_2 ~=~q^{-1/8}\,(1+...)~. \cr }
\eqn\pbfspliteight
$$
Once again $A_8^b-A_8^f=A_8$.  Then our $[8,2]$ identities,
which contain the $[8,0]$ Jacobi-like identities $A_8=B_8=C_8=0$ as a subset,
are as follows:
$$  \eqalign{
     A_8^b ~&=~ \thetatwo/\eta^3 \cr
     A_8^f ~&=~ \thetatwo/\eta^3 \cr
     B_8~&=~C_8 ~=~ 0 \cr
     E_8 ~&=~ \half\,(\thetathree-\thetafour)/\eta^3 \cr
     F_8 ~&=~ \half\,(\thetathree+\thetafour)/\eta^3~. \cr}
\eqn\pbfspliteightids
$$
As a consequence of \pbfspliteightids, we have
$$  (E_8+F_8)^4 ~-~ (E_8-F_8)^4 ~-~ 16\,(A_8^b+A_8^f)^4 ~=~ \deltaminushalf\,J
    ~=~ 0 ~.
\eqn\pbfspliteightcorr
$$
Similarly, for $K=16$, we define the five quantities:
$$   \eqalign{
    A_{16}^b~&\equiv~ d^2_0 ~-~\half\,d^8_0 \cr
    A_{16}^f~&\equiv~ \half\,d^8_8 ~-~d^{14}_8 \cr
    C_{16}~&\equiv~ d^{14}_4 ~-~ \half\,d^8_4 \cr
    E_{16}~&\equiv~ \half\,d^8_6~-~ d^{14}_6~=~q^{7/16}(1+...)\cr
    F_{16}~&\equiv~ d^{2}_2 ~-~ \half\,d^8_2~=~q^{-1/16}(1+...)~;\cr  }
\eqn\psixteenthermodefs
$$
here too $A_{16}^b-A_{16}^f=A_{16}$.  Then our $[16,2]$ identities,
which likewise contain the $[16,0]$
Jacobi-like identities $A_{16}=C_{16}=0$ as a subset, are as follows:
$$   \eqalign{
       A_{16}^b ~&=~\sqrt{\thetatwo/(2\eta^3)}~=~      c^1_1 \cr
       A_{16}^f ~&=~\sqrt{\thetatwo/(2\eta^3)}~=~      c^1_1 \cr
       C_{16} ~&=~ 0 \cr
       E_{16} ~&=~ \half\left( \sqrt{\thetathree/\eta^3} \,-\,
              \sqrt{\thetafour/\eta^3} \right) ~=~ c^2_0 \cr
       F_{16} ~&=~ \half\left( \sqrt{\thetathree/\eta^3} \,+\,
              \sqrt{\thetafour/\eta^3} \right) ~=~ c^0_0 ~\cr}
\eqn\ksixteenthermoids
$$
where {\it on the right sides of these equations the string functions are
at level}\/ $K=2$.  As a consequence of \ksixteenthermoids, we have
$$   (E_{16}+F_{16})^8 - (E_{16}-F_{16})^8 - {1\over{16}}(A_{16}^b+A_{16}^f)^8
      ~=~ \deltaminushalf\,J ~=~0.
\eqn\ksixteencorollary
$$

The identities \ksixteenthermoids\ are in fact quite remarkable,
for they are linear relations indicating that
the differences of certain $K=16$ string
functions are nothing but the $K=2$ string functions.
The full consequences of these relations between the $\bZ_2$ fermionic
characters and the $\bZ_{16}$ parafermionic characters are yet to be
explored.

\section{Third Series: The $[K,1]$ Identities}

This new series of identities generalizes the $\vartheta$-function identity
$$     \vartheta_2 \,\vartheta_3 \,\vartheta_4 ~=~ 2\,\eta^3
\eqn\tobegeneralized
$$
to levels $K>2$, thereby relating the {\it parafermion}\/ characters
$c^\ell_n$ to the boson character $\eta$.

This series of identities actually starts at $K=1$, where we
have the relation \kKzerostringeta:
$$   \eta\,c^0_0 ~=~1~.
\eqn\Kequaloneparabosonization
$$
For $K=2$, as mentioned, we have the identity
$$     \vartheta_2 \,\vartheta_3 \,\vartheta_4 ~=~ 2\,\eta^3~,
\eqn\bosonization
$$
which can be written in terms of the $K=2$ string functions
as
$$   \eta^3 \,Q_2 ~=~ 1
\eqn\Kequaltwoparabosonization
$$
where
$$     Q_2~\equiv~ c^1_1 \,\biggl\lbrack
       (c^0_0)^2~-~(c^2_0)^2 \biggr\rbrack ~=~
     \eta^{-3}\,\sqrt{{{\thetatwo\thetathree\thetafour}\over{2\,\eta^3}}}~.
\eqn\Qtwodef
$$
[Thus \Kequaltwoparabosonization\ is in fact equivalent to
the {\it square root}\/ of \bosonization, and thereby contains the
extra information about the sign of the square root.]
Eqs.~\Kequaloneparabosonization\ and \Kequaltwoparabosonization\
are the first two identities in our $[K,1]$ series, each of
the form $\eta^p \sum (c)^p=1$ for some power $p$.
The corresponding $[K,1]$ identities for $K>2$ are as follows.
For $K=4$, we have
$$  \eta^2\,Q_4 ~=~1
\eqn\Kequalfourparabosonization
$$
where
$$ Q_4 ~\equiv~  (c^0_0 + c^4_0) \,c^2_2
         ~-~ 2\,c^2_0 \, c^4_2 ~,
\eqn\Qfourdef
$$
and for $K=8$, we have
$$   \eta^3 \,Q_8 ~=~1
\eqn\Kequaleightparabosonization
$$
where
$$  \eqalign{
     Q_8 ~\equiv~
     c^4_4\,(c^2_0 + c^6_0)\,\biggl\lbrack
        (c^0_0 + c^8_0) ~-~ (c^0_2 + c^8_2) \biggr\rbrack &\cr
  ~+~ 2\,c^4_2\,\biggl\lbrack
     c^8_4\,(c^2_0+c^6_0) ~-~ c^6_4\,(c^0_0 + c^8_0) \biggr\rbrack &\cr
  ~-~ 2\,c^4_0\,\biggl\lbrack
     c^8_4\,(c^2_2+c^6_2) ~-~ c^6_4\,(c^0_2 + c^8_2) \biggr\rbrack & ~.\cr}
\eqn\Qeightdef
$$
Note that this expression for $Q_8$ can be rewritten in a
variety of forms due to the $[8,0]$ Jacobi-like identities
$A_8=B_8=C_8=0$.  Similarly, for $K=16$,
we define the string-function combinations $d^\ell_n\equiv
c^\ell_n+c^{K-\ell}_n$. Then our corresponding $[16,1]$ identity
is
$$  \eta^3\, Q_{16} ~=~ 1 ~,
\eqn\sixteenform
$$
where $2Q_{16}$ is the following quantity:
$$ \eqalign{
  &~-~2\,d^{0}_{0}\,d^{4}_{2}\,d^{16}_{2}~-~3\,d^{0}_{0}\,d^{6}_{2}\,
  d^{6}_{2}~+~3\,d^{0}_{0}\,d^{6}_{6}\,d^{6}_{6}~+~2\,d^{0}_{0}\,
   d^{12}_{6}\,d^{16}_{6}~-~\phantom{3}\,d^{2}_{0}\,d^{2}_{2}\,d^{2}_{2}\cr
 &~-~4\,d^{2}_{0}\,d^{2}_{2}\,d^{8}_{2}~-~\phantom{3}\,d^{2}_{0}\,
  d^{8}_{2}\,d^{8}_{2}~+~\phantom{3}\,d^{2}_{0}\,d^{8}_{6}\,d^{8}_{6}~
  +~4\,d^{2}_{0}\,d^{8}_{6}\,d^{14}_{6}~+~\phantom{3}\,d^{2}_{0}\,
   d^{14}_{6}\,d^{14}_{6}  \cr
 &~+~2\,d^{4}_{0}\,d^{4}_{2}\,d^{6}_{2}~-~3\,d^{4}_{0}\,d^{16}_{2}
  \,d^{16}_{2}~-~2\,d^{4}_{0}\,d^{6}_{6}\,d^{12}_{6}~+~3\,d^{4}_{0}
   \,d^{16}_{6}\,d^{16}_{6}~+~3\,d^{6}_{0}\,d^{4}_{2}\,d^{4}_{2}  \cr
 &~-~2\,d^{6}_{0}\,d^{6}_{2}\,d^{16}_{2}~+~2\,d^{6}_{0}\,d^{6}_{6}\,
    d^{16}_{6}~-~3\,d^{6}_{0}\,d^{12}_{6}\,d^{12}_{6}~-~\phantom{3}\,
   d^{8}_{0}\,d^{2}_{2}\,d^{2}_{2}~+~\phantom{3}\,d^{8}_{0}\,d^{14}_{6}
     \,d^{14}_{6}  ~.\cr
      }
\eqn\Qsixteendef
$$
This expression for $Q_{16}$ can also be rewritten in many different forms
by using the $[16,0]$ Jacobi-like identities $A_{16}=C_{16}=0$.
Thus, defining $Q_1\equiv c^0_0=\eta^{-1}$ for $K=1$,
we have
$$    {Q_1}^3~=~ Q_2~=~Q_1\,Q_4 ~=~Q_8~=~Q_{16}~.
\eqn\moreids
$$

\section{Comments On Other Identities}

As indicated in the Introduction, these $[K,0]$, $[K,1]$, and $[K,2]$
identities are undoubtedly only some of the general
$[K_1,K_2]$ identities which exist.
Of course, implicit in the above identities are $[K_1,K_2]$
relations which do {\it not}\/ involve the $K=1$ or $K=2$ string functions;
for example, we have the identities \pbfcorollary,
\pbfspliteightcorr, \ksixteencorollary, and \moreids,
as well as additional identities such as
$$   \eqalign{
    \pm(C_4\pm D_4) ~&=~ (E_8\pm F_8)^2 ~=~ (E_{16}\pm F_{16})^4~, \cr
    {\textstyle{1\over 4}} \,(A_4^b+A_4^f)^2 ~&=~
      (C_8+D_8)^4 ~+~ (E_{16}-F_{16})^8 ~,\cr
    ({Q_{16}})^6 ~&=~\half\, (A_8^b+A_8^f)\,({F_8}^2-{E_8}^2)\,({Q_4})^6~. \cr}
\eqn\moreidstwo
$$
However, these identities are not independent of those in the
three preceding series.

We also point out that there can exist {\it several}\/ distinct series
of a given $[K_1,K_2]$ type.  For example,
although the above three series are all of the form
$$   \sum \,(c_{K=K_1})^p ~=~ \sum \,(c_{K=K_2})^p
\eqn\kgenform
$$
where $p$ is an arbitrary power, there also exist more general
``mixed'' identities of the form
$$    \sum \,\biggl\lbrace (c_{K=K_1})^p \,(c_{K=K_2})^q \biggr\rbrace
  ~=~   \sum \,\biggl\lbrace (c_{K=K_1})^p \,(c_{K=K_2})^q \biggr\rbrace~
\eqn\kgenformtwo
$$
which cannot be rewritten in the form \kgenform\ and which
are indeed independent of all such identities.
For example, consider the following three mixed identities involving
the $K=2$ and $K=4$ string functions (where we have distinguished
the two sets of string functions by writing the $K=2$ string
functions in terms of the Jacobi $\vartheta$-functions):
$$  \eqalign{
  A_4^b~ (\thetathree^2+\thetafour^2) ~&=~
          8\,D_4\,\thetatwo^2~,\cr
            \biggl[(d^2_2)^4 \,+\,4\,d^2_2 (d^4_2)^3\biggr] \,
             (\thetathree^2+\thetafour^2) ~&=~
    \biggl[(d^2_0)^3 d^2_2 \,+\,4\,(d^0_0)^3 d^2_2 \biggr]\,
        \thetatwo^2 ~,\cr
   \biggl[4\,(d^0_0)^2 d^2_2 d^4_2 \,-\,2\,(d^2_0)^2 (d^2_2)^2 \biggr]\,
             (\thetathree^2-\thetafour^2) ~&=~
         \biggl[-4\,(d^0_0)^2 d^2_0 d^4_2 \,-\, (d^2_0)^3 d^2_2 \biggr]\,
   \thetatwo^2~.\cr}
\eqn\mixedidentities
$$
While the first of these identities follows directly from
\pbfsplitfour, the remaining two are in fact {\it new}\/ identities
independent of any presented thus far.

It thus appears that there are many unexpected identities involving
string functions at different levels $K$,
implying a rich set of relations between the characters
of different $\bZ_K$ parafermions (and suggesting
various as-yet-undiscovered relationships between the different
$\bZ_K$ parafermion theories themselves).
In Sect.~4 we will prove the identities in our three series,
and discuss how others, such as those in \mixedidentities, might
be obtained.

\endpage

\chapter{PROOFS OF THE IDENTITIES}

In this section we prove the series of identities listed in Sect.~3.
We will find that the proofs of the new $K>2$ cases exactly mirror
the traditional proofs for the known $K=2$ special cases, thus demonstrating
that each series of identities shares the same underlying mathematical
basis.  Our proofs make use of some fundamental and powerful results from
the theory of modular functions, suitably generalized so as to be appropriate
for the $K>2$ cases.
In the first part of this section, therefore, we provide a review
of these results from modular function theory, ultimately quoting
a theorem upon which our proofs rest.
The second part of this section then contains the proofs of
our identities.
We have kept our discussion sufficiently general throughout
in the hope that methods of obtaining and proving additional
$[K_1,K_2]$ identities will become self-evident.

\section{Results from Modular Function Theory}

We first provide a review of those aspects of modular function theory
which will be relevant for the proofs of our identities.  For more
details, we refer the reader to any of the standard modular function
theory references;\REFS\koblitz{
  N. Koblitz, {\it Introduction to Elliptic
  Curves and Modular Forms} (Springer-Verlag, New York, 1984).}
  \REFSCON\gunning{R.C. Gunning, {\it Lectures on Modular Forms}
  (Princeton University Press, Princeton, 1962).}
  \REFSCON\langetal{S. Lang, {\it Introduction to Modular Forms}
  (Springer-Verlag, New York, 1976);
  J. Lehner, {\it Lectures on Modular Forms}, National Bureau of Standards
  Applied Mathematics Series No. 61 (U.S. GPO, Washington, D.C., 1969);
  R.A. Rankin, {\it Modular Forms and Functions}
  (Cambridge University Press, Cambridge, England, 1977);
  J.-P. Serre, {\it A Course in Arithmetic} (Springer-Verlag,
  New York, 1973).}\refsend\
in particular, our approach is based upon those of Refs.~[\koblitz]
and [\gunning].

The homogeneous modular group is $\tilde \Gamma\equiv SL(2,\bZ)$, the
group formed by the set of $2\times 2$ matrices with integer entries and
unit determinant under matrix multiplication.
The inhomogeneous modular group $\Gamma$ (the so-called
``modular group'') is the quotient group $\Gamma \equiv PSL(2,\bZ) \equiv
\tilde \Gamma/\lbrace \pm \bone \rbrace$, the subgroup of $\tilde \Gamma$
in which every matrix $A\in \tilde \Gamma$
is identified with the matrix $-A$.
We shall need to consider various subgroups of $\Gamma$.
Three series of subgroups of $\Gamma$ may be defined as follows:
$$   \eqalign{
    \Gam(N) ~&:~~~~ \left\lbrace A\,\equiv \,\smallmatrix{a}{b}{c}{d}
     \,\in\Gamma ~|~
           a\,\equalmod{N}\,d\,\equalmod{N}\,1,~~
           b\,\equalmod{N}\,c\,\equalmod{N}\,0 ~\right\rbrace~,\cr
    \Gam_1(N) ~&:~~~~ \left\lbrace A\,\equiv \,\smallmatrix{a}{b}{c}{d}
     \,\in\Gamma ~|~
           a\,\equalmod{N}\,d\,\equalmod{N}\,1,~~
           c\,\equalmod{N}\,0 ~\right\rbrace~,\cr
    \Gam_0(N) ~&:~~~~ \left\lbrace A\,\equiv \,\smallmatrix{a}{b}{c}{d}
     \,\in\Gamma ~|~
           c\,\equalmod{N}\,0 ~\right\rbrace~,\cr}
\eqn\ksubgroups
$$
where $\equalmod{N}$ signifies equality modulo $N$.
We thus see that the elements of $\Gam(N)$, $\Gam_1(N)$, and $\Gam_0(N)$
are respectively of the forms $\smallmatrix{1}{0}{0}{1}$,
$\smallmatrix{1}{\ast}{0}{1}$, and $\smallmatrix{\ast}{\ast}{0}{\ast}$
modulo $N$ (where the asterisk indicates the absence of any
defining relation), and therefore
$\Gam(N) \subset \Gamma_1(N)\subset \Gamma_0(N) \subset \Gamma$
for $N>1$.  For $N=1$ these groups each equal $\Gam$.
Such groups are called {\it congruence subgroups} of $\Gamma$,
with $\Gam(N)$ called the {\it principal}\/ congruence subgroups;
in general a congruence
subgroup $\Gam'\subset \Gam$ is said to be of {\it level}\/ $N$
if $\Gam(N)\subset \Gam'$.  (This level $N$ bears no relation to
the Ka\v{c}-Moody level $K$.)
Only the principal congruence subgroups $\Gamma(N)$ are normal
subgroups of $\Gamma$, and in fact
this series of subgroups will concern us the most.
Note that $\Gam(N)$
is isomorphic to $\Gam/SL(2,\bZ_N)$,
where $\bZ_N$ is the set of integers modulo $N$.
In particular, $SL(2,\bZ_1)\approx \lbrace\bone\rbrace$ since all
elements of $SL(2,\bZ)$ are, modulo 1, isomorphic to the identity.

Each of the congruence subgroups $\Gam'$ in \ksubgroups\ has finite index
in $\Gam$:
$$   [\Gam:\Gam'] ~\equiv~{\rm dim}\,\Gam/\Gam' ~<~\infty~,
\eqn\kfiniteindex
$$
and straightforward number-theoretic arguments\refmark{\koblitz,\gunning}\
yield the values:
$$  \eqalign{
    [\Gam:\Gam(N)]~&=~ \epsilon_N\,N^3 \,\prod_{p|N}\,(1-p^{-2})~\cr
    [\Gam:\Gam_1(N)]~&=~ \epsilon_N\,N^2 \,\prod_{p|N}\,(1-p^{-2})~\cr
    [\Gam:\Gam_0(N)]~&=~ \phantom{\epsilon_N}\,
       N\phantom{^2} \,\prod_{p|N}\,(1+p^{-1})~\cr}
\eqn\kindices
$$
where the products are taken over all primes $p>1$ dividing $N$ and where
$$  \epsilon_N ~\equiv~\cases{
        1 & for $N=1,2$ \cr
        1/2 & for $N>2$. \cr}
\eqn\kepsilondef
$$
These factors of $\epsilon_N$
in \kindices\ reflect the fact that $\bone\,\equalmod{N}\,-\bone$
for $N=1,2$, but not for $N>2$.
 From \kindices, therefore, we find
$$   [\Gam:\Gam(N)]~=~ \cases{
             1 & for $N=1$ \cr
             6 & for $N=2$ \cr
             24 & for $N=4$. \cr}
\eqn\kindexcases
$$

For any congruence subgroups $\Gam'$ with finite index in $\Gam$, we can
identify two generators $\gamma_i\in \Gam'$ ($i=1,2$) and a set of
coset representatives $\alpha_j\in \Gam$ ($j=1,...,[\Gam:\Gam']$).
The $\gamma_i$ are generators in the sense that every element of $\Gam'$
can be written as a ``word'' in $\gamma_1$ and $\gamma_2$.
The set of coset representatives (also called a {\it transversal}\/)
contains one element from each coset of $\Gam/\Gam'$;
therefore only one of these coset representatives is in $\Gam'$ itself,
and we are free to choose this representative to be $\alpha_1=\bone$.
At level $N=1$, we have only the full modular group $\Gam$:
its two generators are
$$   S~\equiv~\pmatrix{0 & -1 \cr 1 & 0 }~~~~~~{\rm and}~~~~~~
     T~\equiv~\pmatrix{1 & 1 \cr 0 & 1 }~,
\eqn\kSTdef
$$
and its sole ``coset'' representative is of course $\alpha_1=\bone$.
At higher levels there are more possibilities.
At levels $N=2$ and $N=4$, for example, we find the following
generators and transversals for the principal congruence subgroups:
$$  \eqalign{
     \Gam(2):&~~~~{\rm generators:}~~T^2, ST^{-2}S    \cr
             &~~~~{\rm transversal:}~~ \lbrace\bone, S, T, ST, TS, T^{-1}ST
                                       \rbrace \cr
     \Gam(4):&~~~~{\rm generators:}~~T^4, ST^{4}S    \cr
             &~~~~{\rm transversal:}~~\lbrace X, T^2X, ST^2SX, T^2ST^2SX
                                       \rbrace\cr
             &~~~~\phantom{\rm transversal:}{\rm where}~
                   X\,\equiv\,\lbrace \bone,S,T,ST,TS, TST\rbrace ~,\cr}
\eqn\kgentransvlist
$$
whereas for the $\Gam_0$ subgroups we find:\foot{
   These are actually {\it right}\/ transversals (representatives
   of {\it right}\/ cosets).  For the $\Gam(N)$ subgroups the right
   and left transversals coincide because $\Gam(N)$ is a normal
   subgroup of $\Gam$ for every $N$.}
$$  \eqalign{
     \Gam_0(2):&~~~~{\rm generators:}~~T, ST^{2}S    \cr
             &~~~~{\rm transversal:}~~ \lbrace \bone, S, ST \rbrace \cr
     \Gam_0(4):&~~~~{\rm generators:}~~T, ST^{-4}S    \cr
             &~~~~{\rm transversal:}~~ \lbrace \bone, S,ST,ST^2,ST^3,ST^2S
                                      \rbrace ~.\cr}
\eqn\kotherlist
$$
Note that $\Gamma_0(N)$ and $\Gamma_1(N)$ contain $T$ for all $N$;
therefore one of the generators of these groups is
always $T$.  This is not the case
for the principal subgroups $\Gamma(N)$, which contain $T$ only for $N=1$.
As we shall see, we will be primarily interested in those subgroups
for which $T$ is {\it not}\/ a generator.

The modular group $\Gamma$ is isomorphic to the set of linear-fractional
transformations
$$   \tau ~\to~ {{a\tau+b}\over{c\tau+d}}~,~~~ad-bc~=~1~, ~~~a,b,c,d\in\bZ~,
\eqn\klinfrac
$$
of $\tau\in H$ (where $H$ is the complex upper half-plane);  indeed,
we can identify the transformation \klinfrac\ with the group element
$\smallmatrix{a}{b}{c}{d}\in \Gam$.
A fundamental domain $\calF\equiv\calF[\Gam]$ for $\Gamma$, therefore,
is a set of points $\tau\in H$ such that no two are related by
a transformation of the form \klinfrac.
It is conventional to choose this domain to be contiguous and symmetric
about the $\tau_2$ axis (where $\tau_1$ and $\tau_2$ are respectively
the real and imaginary parts of $\tau$):
$$   \calF~\equiv~\calF[\Gam]~\equiv\lbrace \tau \in {\bf C}~|~
          \tau_2>0,~|\tau_1|\leq \half,~|\tau|\geq 1~\rbrace~.
\eqn\kfunddomain
$$
A fundamental domain $\calF[\Gam']$ corresponding to any {\it subgroup}\/
$\Gam'\subset \Gam$ must therefore be {\it larger}\/ than $\calF[\Gam]$,
and is in fact the totality of points
obtained by acting upon each point in $\calF[\Gam]$ with each of
the coset representatives (including $\bone$) of $\Gam'$:
$$  \calF[\Gam']~\equiv~ \bigcup_{j=1}^{[\Gam:\Gam']}\,\alpha_j\,\calF[\Gam]~.
\eqn\kfunddomainsub
$$
One typically chooses the transversal $\lbrace\alpha_j\rbrace$ in such a way
that with the choice \kfunddomain,
the domain $\calF[\Gam']$ in \kfunddomainsub\ is contiguous.

There are certain points in the complex upper
half-plane $H$ which are called {\it cusp points}:  these are
the point $\tau_\infty \equiv (0,\infty)\equiv i\infty$, along with
the set ${\bf Q}$ (\ie, the points with $\tau_2=0$ and rational
values of $\tau_1$).
We shall need to distinguish those cusp points which
are $\Gam'$-inequivalent for a given congruence subgroup $\Gam'\subset\Gam$
(\ie, those cusp points not related by a transformation in $\Gam'$).
At level $N=1$ the only congruence subgroup is $\Gam$ itself,
and indeed it is a simple matter to see
that all of the cusp points are in the same $\Gam$-equivalence
class.  The set of $\Gam$-inequivalent cusp points therefore contains
only one element, and the typical choice $\tau_\infty$ is consistent
with the choice of fundamental domain \kfunddomain.
Given this choice, it follows that at any higher level $N>1$,
all $\Gam'$-inequivalent cusps for $\Gam'\subset \Gam$ must
be in the set $\lbrace \alpha_i \tau_\infty\rbrace $
where $\lbrace \alpha_i, i=1,...,[\Gam:\Gam']\rbrace$ is the transversal
of $\Gam'$ in $\Gam$.
Such points $\alpha_i\tau_\infty$ for $i>1$ are all $\in {\bf Q}$.
However, even the points in this restricted set are not
necessarily $\Gam'$-inequivalent:  in general two such points
$\alpha_i \tau_\infty$ and $\alpha_j \tau_\infty$ are $\Gam'$-inequivalent
if and only if there does not exist an integer $n$ such
that $\alpha_j T^n \alpha_i^{-1}\in \Gam'$.
It turns out, for example, that for $\Gam'=\Gam_0(p)$ with $p$ a prime number
there are only {\it two} $\Gam'$-inequivalent cusp points
($0$ and $\tau_\infty$), while for $\Gam'=\Gam_0(p^2)$ there are
$p+1$ such cusps points:  $0$, $\tau_\infty$, and $-1/(rp)$ for $r=1,...,p-1$.
A listing of the $\Gam'$-inequivalent cusp points for the congruence
subgroups at levels $N=2$ and $N=4$ which will be relevant to our
later discussion is as follows:
$$ \eqalign{
      \Gam_0(2):&~~~~~{\rm cusps:~~}   0,\,\tau_\infty \cr
      \Gam_0(4):&~~~~~{\rm cusps:~~}   0,\,\tau_\infty,\,-1/2 \cr
      \Gam(2):&~~~~~{\rm cusps:~~}   0,\,\tau_\infty,\,-1 \cr
      \Gam(4):&~~~~~{\rm cusps:~~}   0,\,\tau_\infty,\,
         -1,\,-1/2,\,1,\,3/2,\,2 ~.\cr }
\eqn\kcusps
$$
Such points are called cusp points of $\Gam'$ because
in each case fundamental domains $\calF[\Gam']$
can be chosen whose shapes are cusp-like at each of these points
(and ``cusp''-like at $\tau_\infty$).

We can now rigorously define modular functions, modular forms,
and cusp forms with respect to these general groups $\Gam'\subseteq\Gam$.
First we consider the full modular group $\Gam$ (\ie, level $N=1$).
A {\it modular function} of weight $k\in 2\bZ$ with respect to $\Gam$ is
defined
to be a function $f(\tau)$ satisfying two conditions.  First, it must
have an expansion in powers of $q\equiv \exp\lbrace 2\pi i\tau \rbrace$
of the form
$$      f(\tau) ~=~ \sum_{n\in\bZ}\,a_n\,q^n
\eqn\kqexpansion
$$
where there exists an $m\in\bZ$ such that $a_n=0$ for
all $n<m$ (\ie, there can be at most finitely many non-zero values
of $a_n$ with $n<0$).  This condition therefore ensures that $f(\tau)$
is meromorphic at $\tau= \tau_\infty$ (\ie, $q= 0$).
Second, $f(\tau)$ must satisfy
$$     f(\gamma\tau) ~=~ (c\tau+d)^k \,f(\tau)
\eqn\kmodtrans
$$
for all $\gamma\equiv\smallmatrix{a}{b}{c}{d} \in\Gam$ and $\tau\in H$.
It is convenient to define the {\it stroke
operator} $[\alpha]$ for any $\alpha\equiv
\smallmatrix{a}{b}{c}{d} \in SL(2,\bZ)$:
this operator $[\alpha]$ transforms a modular function $f$ of weight $k$
to $f[\alpha]$, where
$$ (f[\alpha])(\tau)~\equiv~ (c\tau+d)^{-k} \,f(\alpha \tau)~.
\eqn\kstrokedef
$$
With this notation, then, \kmodtrans\ becomes the requirement that
$$    f[\gamma] ~=~ f~~~~~~~~~~ {\rm for~all}~\gamma\in\Gam~,
\eqn\kmodtransstroke
$$
and it is clear that any function $f$ invariant under the two
$\Gam$-generators $[S]$ and $[T]$ therefore satisfies \kmodtransstroke.
Note that the set of modular functions of a given weight $k$
forms a complex vector space, and that the product
of two modular functions of weights $k_1$ and $k_2$ respectively
is a modular function of weight $k_1+k_2$.

Any such modular function $f$ which
additionally satisfies $a_n=0$ for all $n<0$ in \kqexpansion\
is said to be a modular {\it form} with respect to $\Gam$.
Modular forms are therefore holomorphic (rather than merely meromorphic)
at $\tau = \tau_\infty$, remaining finite at $q= 0$.
If in fact $a_0=0$ as well, so that the modular form $f$ actually vanishes
at $q=0$, then $f$ is called a {\it cusp form}.
The complex
vector spaces of modular forms and cusp forms of weight $k$
with respect to $\Gam$ are respectively denoted $M_k[\Gam]$ and $S_k[\Gam]$.

It is straightforward to generalize these definitions to congruence
subgroups at higher levels $N>1$.
For any such subgroup $\Gam'$, a modular function $f$ of weight $k\in 2\bZ$
with respect to $\Gam'$ must again satisfy two conditions.  The
natural generalization of \kmodtransstroke\ is the requirement that
$$    f[\gamma] ~=~ f~~~~~~~~~~ {\rm for~all}~\gamma\in\Gam'~,
\eqn\kmodtranssub
$$
and once again it is sufficient to demonstrate that $f[\gamma_i]=f$
for the two generators $\gamma_i$ of $\Gam'$ in order to
demonstrate \kmodtranssub.
The generalization of \kqexpansion, on the other hand, is a bit more
subtle.  Eq.~\kqexpansion\ was the requirement that $f$ be meromorphic
at $q=0$ (\ie, at $\tau=\tau_\infty$, the cusp point of $\Gam$).
For subgroups $\Gam'$ at higher levels $N>1$,
we therefore analogously require that $f$ be meromorphic
at {\it each} of the $\Gam'$-inequivalent cusp points of $\Gam'$.
This meromorphicity is not determined by evaluating
$f(\tau)$ as $\tau$ approaches each of these cusp points, however;
in fact, since each of these additional cusp points is $\in {\bf Q}$,
we have $|q|=1$ at these points and therefore
a straightforward $q$-expansion of $f$ does not converge.
Instead, meromorphicity at the cusps is defined as follows.
We have seen that this set of $\Gam'$-inequivalent
cusp points can be written as $\lbrace \alpha_i \tau_\infty\rbrace$ where
these $\alpha_i$ are among (but not necessarily all of)
the coset representatives of $\Gam'$ in $\Gam$.  Let $s$ denote
the set of these $\alpha_i$ (hence $s$ is a subset of the transversal).
The analogue of \kqexpansion\ is therefore the requirement that
for each $\alpha_i\in s$, we can perform a $q$-expansion
$$   f[\alpha_i]~=~ \sum_{n\in \bZ/N} \,a_n\, q^n
\eqn\kqexpansionsub
$$
where, as before, there exists an $m\in \bZ/N$ such that $a_n=0$
for all $n<m$.
Note that $n$ and $m$ can now take values in the larger set $\bZ/N$
(rather than $\bZ$ itself).  Also note that for $i>1$ we have $\alpha_i
\not\in \Gam'$, so \kqexpansionsub\ is in general quite stringent.

It is clear that at level $N=1$, \kqexpansionsub\ reduces to \kqexpansion,
for in this case $s=\lbrace \alpha_1 \rbrace =\lbrace \bone \rbrace$.
For $N>1$, the $i=1$ case of \kqexpansionsub\ implies meromorphicity
at $\tau_\infty$, and the $i>1$ cases imply meromorphicity at each of
the remaining $\Gam'$-inequivalent cusps $\in {\bf Q}$.  Note that
\kqexpansionsub\ is in fact sufficient to imply that $f$ is actually
meromorphic at {\it all}\/ of the cusp points of $\Gam'$.

Once again, if indeed $a_n=0$ for all $n<0$ and each $\alpha_i\in s$
(so that $f$ is holomorphic at each $\Gam'$-cusp, remaining
finite), then $f$
is deemed a modular {\it form} of weight $k$ with respect to $\Gam'$.
If additionally $a_0=0$ for each $\alpha_i\in s$ (so that $f$ vanishes
at each cusp), then $f$ is a {\it cusp form} with respect to $\Gam'$.
The complex vector spaces of such weight-$k$ modular forms and cusp
forms are denoted
$M_k[\Gam']$ and $S_k[\Gam']$ respectively;  note that
$M_k[\Gam]\subseteq M_k[\Gam']$ for all $\Gam'\subseteq \Gam$,
as well as the property $M_{k_1}[\Gam']M_{k_2}[\Gam']\subseteq
M_{k_1+k_2}[\Gam']$.

We are now in a position to state the fundamental
theorem\refmark{\gunning}\ upon which the proofs of
our identities rest.

\noindent{\underbar{\bf Theorem:}}~~
Let $\Gam'\subseteq\Gam$ be any level-$N\geq 1$
congruence subgroup of the modular group $\Gam$, and let $M_k[\Gam']$
denote the space of modular forms of weight $k\in 2\bZ$ with respect
to $\Gam'$.  Then the sizes of such spaces depend on $k$ as follows:
\item{\bullet}  dim $M_k[\Gam']=0$ ~~for all $k<0$.
\item{\bullet}  dim $M_k[\Gam']=1$ ~~for $k=0$.
\item{\bullet}  For $k>0$ a general formula exists as well.
Although we will not require these $k>0$ results for the proofs
of our specific identities, we include the following two special
cases which are likely to be useful in proving additional
$[K_1,K_2]$ identities:
\itemitem{$\bullet$}  For level $N=1$ (\ie, $\Gam'=\Gam$), we have
$$   {\rm dim}~M_k[\Gam]~=~\cases{
    [k/12] & for $k~\equalmod{12}~2$ \cr
    [k/12] +1 & otherwise \cr}
\eqn\ktheoremone
$$
where $[x]$ is the greatest integer $\leq x$, and $\equalmod{12}$
signifies equality modulo $12$.
\itemitem{$\bullet$}  For the principal congruence subgroups $\Gam(N)$
at levels $N>1$, we have instead:
$$  {\rm dim}~M_k[\Gam(N)] ~=~ {{(k-1)N+6}\over{12\,N}} ~ [\Gam:\Gam(N)]
\eqn\ktheoremtwo
$$
where the index $[\Gam:\Gam(N)]$ is given in \kindices.

Thus, for example, in our cases of interest this theorem tells us
that for $\Gam'=\Gam$ we have
$$     {\rm dim}~M_k[\Gam]~=~ \cases{
      0 & for $k<0$ or $k=2$\cr
      1 & for $k=0,14$ and $4\leq k \leq 10$, \cr}
\eqn\kdimsgamma
$$
whereas for $\Gam'=\Gam(N)$ we have
$$  {\rm dim}~M_k[\Gam(N)]~=~ \cases{
    0 & for all $N>1$, $k<0$ \cr
    1 & for all $N>1$, $k=0$ \cr
   \half k +1 & for $N=2$, $k>0$ \cr
   2k+1 & for $N=4$, $k>0$. \cr}
\eqn\kdimsgammaprime
$$
These dimensions are important, for they tell us the number of
``basis'' modular forms in terms of which any modular form
of weight $k$ can be expressed as a polynomial.
For example, since ${\rm dim}~M_0[\Gam']=1$
for all congruence subgroups $\Gam'\subseteq \Gam$,
and since $f=1$ is a valid $\Gam'$-modular form of weight
$k=0$, {\it all}\/ $\Gam'$-modular forms of weight $k=0$ must be constants:
$$  M_0[\Gam'] ~=~ {\bf C} ~
\eqn\kconstants
$$
where ${\bf C}$ is the space of complex numbers.
Similarly, since ${\rm dim}~M_k[\Gam']=0$ for all $k<0$,
all $\Gam'$-modular forms of negative weight must vanish identically:
$$  f~\in~M_k[\Gam']~~\Longrightarrow~~f~=~0~~~~~~{\rm for~all}~k<0~.
\eqn\kvanishings
$$
Likewise, for $k=4,6,8,10,\,{\rm or}\,14$,
we see that all $f\in M_k[\Gam]$ must be multiples of a single
function $E_k$;  these ``basis functions'' $E_k(\tau)$ form the
so-called {\it Eisenstein series}.\refmark{\koblitz-\langetal}\

Finally, we remark that equally powerful results can be obtained for the
spaces of cusp forms $S_k[\Gam']$, as well as for the cases when $k$
is odd (and in fact half-integral).  However, the above results will
be sufficient for the proofs of our series of identities.

\section{Proofs of the Identities}

Given the theorem presented in the last subsection,
it is relatively straightforward
to prove each of the series of identities listed in Sect.~3.
The basic idea of each proof is the same:  we demonstrate
that our string-function expressions are modular forms of a given
weight with respect to an appropriate congruence subgroup
$\Gam'\subseteq \Gam$, whereupon the theorem allows us
to conclude the claimed identity.
The primary subtleties involve properly formulating the identities
and identifying the relevant congruence subgroups.

Since all of our identities involve combinations of the Dedekind
$\eta$-function, the Jacobi $\vartheta_i$-functions, and the string
functions $c^\ell_n$, let us first recall how these functions
transform under the modular group.  Under $T$, their respective
transformation rules take the forms:
$$   \eqalign{
    \eta(\tau+1)~&=~\alpha\,\eta(\tau) \cr
    \vartheta_i(\tau+1)~&=~ \sum_j \,\alpha_{ij}\,\vartheta_j(\tau)\cr
    c^\ell_n(\tau+1) ~&=~ \alpha'\,c^\ell_n(\tau) \cr }
\eqn\ktbehavior
$$
(where $\alpha$, $\alpha'$, and $\alpha_{ij}$ indicate various phases
and mixing matrices), while under $S$ we have
$$   \eqalign{
    \eta(-1/\tau)~&=~\sqrt{\tau}\,\beta\,\eta(\tau) \cr
    \vartheta_i(-1/\tau)~&=~\sqrt{\tau}\,
            \sum_j \,\beta_{ij}\,\vartheta_j(\tau)\cr
    c^\ell_n(-1/\tau) ~&=~ \left(\sqrt{\tau}\right)^{-1}
     \sum_{\ell',n'}\,\beta^{\ell \ell'}_{n n'}\,c^\ell_n(\tau) \cr }
\eqn\ksbehavior
$$
(where again the $\beta$'s represent various phases and mixing
matrices).
Thus, we see that even though these functions themselves are not
invariant under $S$ and $T$, they each transform {\it covariantly}\/
under the modular group, filling out (in the case of $\vartheta_i$
and $c^\ell_n$) representations of the modular group with dimensions
greater than 1.  Furthermore, we see from \ktbehavior\ and \ksbehavior\
that the $\eta$-function and $\vartheta_i$-functions transform with
positive modular weight $k=1/2$, while the string functions $c^\ell_n$
transform with negative modular weight $k=-1/2$.

Given these observations, it is straightforward to determine the modular
weights of each of our identities listed in Sect.~3.
For the first series of identities,
we see that each expression $A_K$, $B_K$, or $C_K$ contains $16/K$
string-function factors;  indeed,
each of these identities takes the general form
$$      \sum  \,(c)^{16/K} ~=~0~.
\eqn\kseriesoneform
$$
Thus, for any Ka\v{c}-Moody level $K$, the level-$K$ identity
in the first series has modular weight $k=-8/K$.
Similarly, the second series of identities involves
combinations of all three of our functions ($\eta$, $\vartheta$, and $c$);
however, for each level $K$, the level-$K$ identity always
takes the general form
$$   \sum\, (c)^{16/K} ~+~\sum \,
     \left( {{\vartheta}\over{\eta^3}} \right)^{8/K} ~=~0~,
\eqn\kseriestwoform
$$
from which it follows once again that the level-$K$ identity in this
series has modular weight $k=-8/K$.
A similar situation exists for the third series as well:  each of these
identities takes the general form
$$    \eta^p\,\sum (c)^p ~=~1
\eqn\kseriesthreeform
$$
for some integer power $p$,
and therefore for any $K$ these identities have modular weight $k=0$.

Hence it is clear that all of these identities can be made to
follow from the theorem given in the last subsection,
provided each can be rewritten in such a manner that their
left sides are modular {\it forms}
of appropriate weight $k$ with respect to a congruence
subgroup $\Gam'\subseteq \Gam$.

 {\bf \underbar{First series -- the $[K,0]$ identities:}}~~

We begin by proving the $K=2$ identity $A_2=0$, as given in Sect.~3.
This is of course the famous Jacobi $\vartheta$-function identity,
and our subsequent proofs will be generalizations of this proof.
Recall that $A_2$ can be rewritten in terms of $\vartheta$ and $\eta$
functions as $\half \Delta^{-1/2} J$ where $\Delta=\eta^{24}$
and $J\equiv \thetathree^4-\thetatwo^4-\thetafour^4$.  Thus it
is clear that $A_2$ has modular weight $k=-8/K=-4$,
and since $A_2[S]=A_2[T]=A_2$,
we see that $A_2$ is a modular function with respect to the full modular
group $\Gam$.  It is also straightforward to check that $A_2$
is {\it tachyon-free}:  by this we mean that in a $q$-expansion
$$   A_2 ~=~ \sum_n \,a_n \,q^n
\eqn\ktachfree
$$
we find $a_n=0$ for all $n<0$.
It then follows that $A_2 \in M_{-4}[\Gam]$, whereupon the theorem
in the last subsection gives the result $A_2=0$.
Note that
since $A_2=\half \deltaminushalf J$, and since it is easy to check
that $\deltaminushalf\not= 0$, this result implies the Jacobi identity
$J=0$.  Note that it would have been more difficult to prove this latter
identity directly, for even though the dimension of $M_2[\Gam]$ is zero,
$J$ itself is {\it not}\/ invariant under $S$ and $T$ (indeed, one
finds $J[S]=J[T]=-J$).
The extra factor of $\deltaminushalf$, which
appears naturally in the definition of $A_2$, absorbs this unwanted
minus sign and leads to the simple proof $A_2=0$.

Let us now proceed to the $K=4$ case:  we wish to prove $A_4=B_4=0$.
Unlike the $K=2$ case, $A_4$ and $B_4$ are not each
invariant under the stroke operators $[S]$ and $[T]$;
rather, they together fill out a {\it two}\/-dimensional representation
of $\Gam$:
$$ \eqalign{
    \pmatrix{ A_4 \cr B_4 \cr}\,[S] ~&=~
      e^{i\pi}\,\pmatrix{
      1/2 & 3 \cr   1/4 & -1/2 \cr} \,\pmatrix{A_4 \cr B_4 \cr} ~\cr
    \pmatrix{ A_4 \cr B_4 \cr}\,[T] ~&=~
       \pmatrix{ 1 & 0 \cr 0 & -1 \cr }
    \,\pmatrix{ A_4 \cr B_4 \cr} ~,\cr}
\eqn\kmixingsfour
$$
and we wish to prove that $A_4$ and $B_4$ {\it individually}\/ vanish.
This is the reason it is necessary to consider the congruence subgroups
$\Gam'\subset\Gam$.

Let us first prove $A_4=0$.  It is clear that $A_4$ is not modular-invariant
under $[S]$, so the full modular group $\Gam$ cannot be the relevant
group in this case.  Instead, let us consider $\Gam_0(2)$.  From
the mixing matrices \kmixingsfour\ we have $A_4[T]=A_4$, as well as
$$   \eqalign{
       A_4[ST^2S]~
      &=~ e^{i\pi}\,\left( \half\,A_4 ~+~ 3\,B_4\right) \,[T^2S] \cr
      &=~ e^{i\pi}\,\left( \half\,A_4 ~+~ 3\,B_4\right) \,[S] \cr
      &=~ (e^{i\pi})^2 \,A_4 ~=~A_4~. \cr }
\eqn\kstepone
$$
The second equality follows from the fact that $A_4$ and $B_4$ are
both invariant under $T^2$;  note also that in general $f[\alpha\beta]=
(f[\alpha])[\beta]$.
Thus, the first condition \kmodtranssub\ for $A_4$ to be a $\Gam_0(2)$
modular form is satisfied: $A_4$ is invariant under all $\gamma\in
\Gam_0(2)$.  We now must show that the second condition
\kqexpansionsub\ is satisfied as well.
We see from \kcusps\ that for $\Gam'=\Gam_0(2)$ there are only two
$\Gam'$-independent cusp points, $\tau_\infty$ and $0$, and therefore
the transversal subset $s$ is only $\lbrace \bone,S\rbrace$.
It is clear that $A_4[\bone]=
A_4$ has a $q$-expansion of the proper form \kqexpansionsub\  (with $m=0$),
and similarly we see that
$A_4[S]=e^{i\pi}(\half A_4+3B_4)$ also has a $q$-expansion
of the proper form (with $m=0$, since {\it both}\/ $A_4$ and $B_4$ are
tachyon-free).  It therefore follows that $A_4\in M_{-2}[\Gam_0(2)]$,
whereupon we obtain the identity $A_4=0$.

Note that there are indeed other ways we might have obtained this
result.  For instance, let us instead consider the congruence subgroup
$\Gam'=\Gam_0(4)$.  It is straightforward to check that $A_4$ is
invariant under both generators $[T^4]$ and $[ST^{-4}S]$,
and for this congruence
subgroup the cusp points are $\tau_\infty$, $0$, and $-1/2$.  We have
already shown that $A_4$ has the proper behavior at the first two
of these cusp points;  let us therefore focus on the third.
Note that $-1/2=\alpha\tau_\infty$ where
$\alpha\equiv \smallmatrix{-1}{\phantom{-}1}{\phantom{-}2}{-3}\in\Gam$.
We therefore must examine $A_4[\alpha]$.  Since it turns out that
$\alpha=ST^2ST^{-1}$, we have
$$  \eqalign{
     A_4\,[ST^2ST^{-1}]~
    &=~ e^{i\pi} \left( \half\,A_4~+~ 3\,B_4\right) \,[T^2ST^{-1}] \cr
       &=~e^{i\pi} \left( \half\,A_4~+~ 3\,B_4\right) \,[ST^{-1}] \cr
       &=~(e^{i\pi})^2 \,A_4\,[T^{-1}] ~=~ A_4~. }
\eqn\ksteptwo
$$
Therefore $A_4$ has the same behavior at the cusp point $-1/2$
as it does at $\tau_\infty$, and since $A_4$ is tachyon-free
(\ie, since $A_4$ remains finite at $\tau_\infty$),
we find $A_4\in M_{-2}[\Gam_0(4)]$.
This again implies the conclusion $A_4=0$.  In fact, one can similarly
demonstrate that $A_4\in M_{-2}[\Gam(2)]$ and $A_4\in M_{-2}[\Gam(4)]$,
each of which leads as well to this result.

Having proven $A_4=0$, we find that there are two ways to prove $B_4=0$.
The first is indirect but simpler:  since
$$  A_4[S]~=~e^{i\pi}\,\left(\half\, A_4~+~3\,B_4\right)~=~0~,
\eqn\ksimpler
$$
we must have $B_4=0$.  A more direct method (not relying on the
identity $A_4=0$) is to construct an independent proof along the above
lines.  Note that since $B_4$ is not invariant under $[T]$, we cannot
consider any subgroups for which $T$ is a generator;  we are therefore
restricted to consideration of the {\it principal}\/ congruence subgroups
$\Gam(N)$.  It is straightforward to demonstrate that $B_4\in M_{-2}[\Gam(N)]$
where $N$ is either $2$ or $4$, and therefore $B_4=0$.
In fact, $B_4$ is a {\it cusp form}\/ with respect to these groups
$\Gam(N)$, since $B_4$ {\it a priori}\/
has a $q$-expansion of the form $q^h(1+...)$ where $h>0$.

Let us now collect together the essential ingredients in these proofs,
in order to frame a general argument.  First, the set of string-function
expressions $\lbrace A_K, B_K,...\rbrace$ must be closed under
$[S]$ and $[T]$, forming a multi-dimensional representation $R_K$
of the modular group $\Gam$;
furthermore, a congruence subgroup $\Gam'\subseteq \Gam$
must be identified such that each member of the representation $R_K$ is
itself {\it invariant}\/ under the generators of $\Gam'$
(\ie, each member must separately comprise a {\it one}-dimensional
representation of $\Gam'$).  Second, this entire representation
$R_K$ must transform under the modular group
with {\it negative} even modular weight $k$.
Third, {\it all}\/ members of $R_K$ must have $q$-expansions
with finitely many non-zero coefficients $a_n$ with $n<0$.
This third condition is needed in order to insure that each
element $\in R_K$ is meromorphic at {\it all}\/ of the cusp points of $\Gam'$,
for any one member of $R_K$ will be meromorphic at {\it all}\/ the
cusp points of $\Gam'$ if and only if {\it all}\/ of the members of $R_K$
are meromorphic at the {\it one}\/ cusp point $\tau_\infty$
(because the set $s$ always contains at least $\bone$ and $S$).
These three conditions then guarantee that
each string-function expression in the representation $R_K$ is itself
a modular function with respect to $\Gam'$.
If each member of the representation is also tachyon-free,
then each is a modular {\it form} with respect to $\Gam'$ and hence
must vanish identically.

Let us now see how to apply this general argument to the
$K=8$ case as we attempt to prove $A_8=B_8=C_8=0$.
The mixing matrices of these expressions under the stroke operators
$[S]$ and $[T]$ are as follows:
$$   \eqalign{
   \pmatrix{ A_8 \cr B_8 \cr C_8 \cr }\,[S]~&=~
      e^{i \pi/2} \,
      \pmatrix{ 1/2 & 1/2 & 1 \cr
                1/2 & 1/2 & -1 \cr
                1/2 &-1/2 & 0 \cr}\, \pmatrix{ A_8 \cr B_8 \cr C_8 \cr } ~\cr
   \pmatrix{ A_8 \cr B_8 \cr C_8 \cr }\,[T]~&=~
      \pmatrix{ 1 & 0 & 0 \cr 0 & -1 & 0 \cr 0 & 0 & -i \cr}\,
                 \pmatrix{ A_8 \cr B_8 \cr C_8 \cr } ~.\cr}
\eqn\ksectfourKeightmixings
$$
There are immediately two problems.  First, we see
that this $K=8$ representation has an {\it odd}\/ modular
weight $k=-8/K=-1$;  our theorem applies only to the cases $k\in 2\bZ$.
Second, no member of this representation is invariant
under {\it any} of the congruence subgroups:  for example, under $\Gam_0(4)$
we find $A_8[T]=A_8$ but
$$    \eqalign{
      A_8[ST^{-4}S] ~
    &=~  e^{i\pi/2}\,\left( \half \,A_8 ~+~ \half \,B_8 ~+~ C_8\right)
                  \,[T^{-4}S] \cr
    &=~  e^{i\pi/2}\,\left( \half \,A_8 ~+~ \half \,B_8 ~+~ C_8\right)
                  \,[S] \cr
    &=~  (e^{i\pi/2})^2\,A_8 ~=~ -A_8 ~.\cr}
\eqn\kstepthree
$$
There are two ways to solve these difficulties.  One possibility
is to extend the theorem presented in Sect.~4.1 to apply to odd $k$
and modular functions with so-called {\it multiplier systems} (\ie, phases
such as the unwanted sign appearing above).  Such extensions can indeed
be made;  in this relatively simple case, for example, we can
instead choose to prove the modified identities
$A_8/\eta^6= B_8/\eta^6=C_8/\eta^6=0$.  These modified
identities would then have modular weight $k= -2$, and the
extra $\eta$-functions absorb the unwanted sign.
A simpler approach, however, (and one which generalizes more
easily to other situations) is to prove instead the
identities $(A_8)^2=(B_8)^2=(C_8)^2=0$, for such identities also have
an even modular weight $k=-2$ and simultaneously avoid such unwanted signs.
(We are essentially enlarging our $K=8$ representation:
$R_8\to R'_8\equiv R_8\otimes R_8$.)
If these quadratic identities can be proven for
all $\tau$, then of course the linear results $A_8=B_8=C_8=0$
immediately follow.
To prove these quadratic identities, we follow the procedure outlined
above:  either choice $\Gam'=\Gam_0(4)$ or $\Gam(4)$ suffices for
proving $(A_8)^2=0$, and $\Gam(4)$ suffices for independently
proving $(B_8)^2=0$ and $(C_8)^2=0$.
It is of course possible to deduce $B_8=C_8=0$ from the result
$A_8=0$ as we did for the $K=4$ case:  in the present case
the analogue of \ksimpler\ is
$$     A_8[S]~=~ e^{i\pi/2} \,\left( \half\,A_8~+~ \half\,B_8~+~ C_8
     \right)~=~0~,
\eqn\ksimplereight
$$
and this implies the weaker result
$B_8+2C_8=0$.  However, $B_8\sim q^{1/2}(1+...)$ and
$C_8\sim q^{3/4}(1+...)$, where inside the parentheses all $q$-exponents
are integral.  Therefore, $B_8$ and $C_8$ must each vanish separately.
Note that in this $K=8$ case we are compelled to consider congruence subgroups
of levels $N\in 4\bZ$ only.  This occurs because the $K=8$ representation
includes a sector $C_8$ with quarter-integer powers of $q$.  Since
in general the generators of $\Gam(N)$ are of the form
$T^N$ and $ST^{\pm N}S$, the choice $N=4$ is the smallest level $N$
for which each element in the $K=8$ representation $R_8$
is $\Gam(N)$-invariant.

The same procedure applies for the $K=16$ case as well;
here we wish to prove $A_{16}=C_{16}=0$, and the relevant mixing matrices
are as follows:
$$ \eqalign{
 \pmatrix{A_{16}\cr C_{16}\cr}~[S]~&=~ e^{i\pi/4}\,
    {1\over{2\sqrt{2}}} \,\pmatrix{ 2 & 4 \cr 1 & -2} \,
   \pmatrix{A_{16}\cr C_{16}\cr}~\cr
 \pmatrix{A_{16}\cr C_{16}\cr}~[T]~&=~
    \pmatrix{1 & 0 \cr 0 & -i \cr}\, \pmatrix{A_{16}\cr C_{16}\cr}~.\cr}
\eqn\pSsixteenagain
$$
Once again we find that we must enlarge our representation
$$    R_{16} ~\to ~ R'_{16} ~\equiv~
      R_{16} ~\otimes~ R_{16} ~\otimes~ R_{16} ~\otimes~ R_{16} ~
\eqn\krepbigger
$$
and prove instead the identities $(A_{16})^4=(C_{16})^4=0$;
similarly, we must choose choose the
level $N=4$ due to the presence of the $C_{16}$ sector in $R_{16}$.
As usual, the subgroup $\Gam(4)$ suffices in general for proving
that each member of $R'_{16}$ vanishes, and we can instead make the
choice $\Gam_0(4)$ in the case of the $A_{16}$ sector (which is
invariant under $T$).  Once again the proof that any one member
of $R_{16}$ vanishes is sufficient to prove that all vanish,
provided each has a different eigenvalue under $T$.
Note that instead of enlarging the representation as in \krepbigger, it
would also have been possible in this case to divide our original
representation by $\eta^3$;  however, this would have necessitated
constructing a proof using congruence subgroups of level $N=8$.

 {\bf \underbar{Second series -- the $[K,2]$ identities:}}~~

The second series of identities is closely related to the first
and in fact contains the first series as a subset.
Recall that for each Ka\v{c}-Moody level $K\in \lbrace 2,4,8,16\rbrace$,
there exists a string-function expression $A_K\sim q^0(1+...)$
which, according to the first series of identities, vanishes identically.
In this second series of identities we show that
the separate bosonic and fermionic pieces of $A_K$ (denoted
$A_K^b$ and $A_K^f$ respectively) can each be written in terms of
Jacobi $\vartheta$-functions;  these
$\vartheta$-function expressions for $A_K^b$ and $A_K^f$
are of course equal, since $A_K=A_K^b-A_K^f=0$.
As a by-product, we also obtain
additional string-function expressions $\lbrace C_K,D_K,...\rbrace$
which can be easily expressed in terms of Jacobi $\vartheta$-functions
as well.

Let us first consider the $K=4$ identities in this series, as given
in \pbfsplitfour:  here we have separated $A_4$ into its separate
bosonic and fermionic contributions $A_4^b$ and $A_4^f$ as
in \pbfsplitfourdefs.
While we know that $A_4$ and $B_4$ together fill out a two-dimensional
representation of the modular group with weight $k=-2$, we find
that the individual pieces $A_4^b$ and $A_4^f$ do not close
separately into only themselves and $B_4$.  Rather, in order to construct
a representation of $\Gam$ containing $A_4^b$ and
$A_4^f$ as separate members, we must introduce the two additional
string-function expressions $C_4$ and $D_4$ defined in \pbfsplitfourdefs.
Together, the set $\lbrace A_4^b, A_4^f, B_4, C_4, D_4 \rbrace$
indeed fills out a complete representation $R_c$ with weight $k=-2$;  each
member of $R_c$ transforms as an eigenfunction under $[T]$:
$$   \pmatrix{ A_4^b \cr A_4^f \cr B_4 \cr C_4\cr D_4 \cr}\,[T]~=~
     \pmatrix{ 1 & 0 & 0 & 0 & 0 \cr
               0 & 1 & 0 & 0 & 0 \cr
               0 & 0 & -1 & 0 & 0 \cr
               0 & 0 & 0 & i & 0 \cr
               0 & 0 & 0 & 0 & -i \cr}\,
   \pmatrix{ A_4^b \cr A_4^f \cr B_4 \cr C_4\cr D_4 \cr}~,
\eqn\ksplitmixTfour
$$
and under $[S]$ they close into each other:
$$   \pmatrix{ A_4^b \cr A_4^f \cr B_4 \cr C_4\cr D_4 \cr}\,[S]~=~
             {{e^{i\pi}}\over 4}\,\pmatrix{
       1 & -1 & 6 & -4 & 4 \cr
      -1 & 1 & -6 & -4 & 4 \cr
       1 & -1 & -2 & 0 & 0 \cr
       -1 & -1 & 0 & 2 &  2 \cr
       1 & 1 & 0 &  2 & 2 \cr} ~
   \pmatrix{ A_4^b \cr A_4^f \cr B_4 \cr C_4\cr D_4 \cr}~.
\eqn\ksplitmixSfour
$$
As a check, note that these matrices indeed contain
the $(A_4,B_4)$ mixing matrices \kmixingsfour\ as the appropriate
submatrices.

Unlike the previous representations we have considered,
not all members of this representation $R_c$ individually
vanish, for while all are modular {\it functions} with respect
to an appropriate $\Gam'\subset \Gam$, not all are modular {\it forms}.
Since the quantity $D_4$ is not tachyon-free, it clearly has the
wrong behavior at the cusp point $\tau=\tau_\infty$;  furthermore,
since the $[S]$-transforms of $A_4^b$, $A_4^f$, $C_4$, and $D_4$
each separately involve $D_4$, each of these quantities has
incorrect (\ie, tachyonic) behavior at the $\Gam'$ cusp point $\tau=0$.
Indeed, only $B_4$ and the difference $A_4=A_4^b-A_4^f$
are free of this tachyonic behavior at both cusp
points $\tau=\tau_\infty$ and $\tau=0$, so of the five quantities in
the above representation $R_c$ only $B_4$ is itself a proper $\Gam'$-modular
form.  Therefore, in order to construct identities for these
five individual quantities,
it is necessary to build a {\it new}\/ representation
involving them in such a manner that {\it each}\/ member is
a tachyon-free modular form.

It turns out that this is not hard to do.  As we have seen in \ktbehavior\
and \ksbehavior, the $\vartheta$- and $\eta$-functions fill out
valid representations of the modular group, and indeed the three quantities
$\lbrace\eta^{-6}\thetatwo^2,\eta^{-6}\thetathree^2,
\eta^{-6}\thetafour^2\rbrace$
fill out such a representation with modular weight $k=-2$.  Let us
take various linear combinations of these quantities, promoting them
to the  ``five''-dimensional representation
$$  R_\vartheta~\equiv~\left\lbrace  {{\thetatwo^2}\over{\eta^6}}\, , ~
      {{\thetatwo^2}\over{\eta^6}}\, , ~
        0\, ,~
      \half\left({{\thetathree^2-\thetafour^2}\over{\eta^6}}\right)\, , ~
      \half\left({{\thetathree^2+\thetafour^2}\over{\eta^6}}\right)
     \right\rbrace~.
\eqn\kRthetadef
$$
Written this way,
this five-dimensional representation
$R_\vartheta$ with weight $k=-2$
has two very important properties.  First, its mixing matrices under
$[S]$ and $[T]$ are (or can be chosen to be) the same as
those in \ksplitmixTfour\ and \ksplitmixSfour\ for $R_c$;
indeed, in this respect the two representations $R_c$ and $R_\vartheta$
transform identically.  More importantly, however,
it is easy to verify that the tachyonic terms within
the fifth member of the $\vartheta$-function representation $R_\vartheta$
are the {\it same} as those of the fifth member $D_4$ of the string-function
representation $R_c$;  additionally, all of the other members of $R_\vartheta$
are themselves tachyon-free.
Therefore, subtracting the two representations, \ie,
$$   R ~\equiv~ R_c~-~R_\vartheta~=~\pmatrix{
       A_4^b ~-~ \thetatwo^2/\eta^6 \cr
       A_4^f ~-~ \thetatwo^2/\eta^6 \cr
                B_4 \cr
         C_4~-~ \half\, (\thetathree^2-\thetafour^2)/\eta^6 \cr
         D_4~-~ \half\, (\thetathree^2+\thetafour^2)/\eta^6 \cr }~,
\eqn\ksubtractedrep
$$
yields a five-dimensional representation $R$\/ in which {\it each}\/
member is tachyon-free.
It is then a straightforward matter to demonstrate that each member
of this new representation $R$
is indeed a modular form with respect to a congruence
subgroup $\Gam'\subset \Gam$
of level $N\in 4\bZ$ (\eg, each member is $\in M_{-2}[\Gam(4)]$),
whereupon it follows that each vanishes identically.
This, then, establishes the identities \pbfsplitfour.

Note that the existence of such identities relies on the existence
of a $\vartheta$-function representation $R_\vartheta$ with the
desired modular weight $k$, the desired transformation matrices $[S]$ and
$[T]$,
and the required tachyonic behaviors of its members.
Such a representation does not always exist.  It is indeed
fortuitous, however, that such representations do exist for
each value of $K$, yielding all of the identities in this second
series.

The derivations of the other identities in this series proceed in
analogous fashion.
For the $K=8$ case, we find that $A_8^b$ and $A_8^f$ are members of
the six-dimensional representation $R_c\equiv\lbrace A_8^b,A_8^f,
B_8,C_8,E_8,F_8\rbrace$ [where these six quantities are defined in
and \ABCKequaleight\ and \pbfspliteight];  under $[S]$ and $[T]$
quantities mix as follows:
$$ \pmatrix{A_8^b \cr A_8^f \cr B_8 \cr C_8 \cr E_8 \cr F_8 \cr}
   \,[S] ~=~ {{e^{i\pi/2}}\over{4}}\,\pmatrix{
       1 & -1 & 1 & 2 & -4 & 4 \cr
      -1 &  1 & -1 & -2 & -4 & 4 \cr
       2 & -2 & 2 & -4 & 0 & 0 \cr
       2 & -2 & -2 & 0 & 0 & 0 \cr
      -1 & -1 & 0 & 0 & 2 & 2 \cr
       1 &  1 & 0 & 0 & 2 & 2 \cr }\,
   \pmatrix{A_8^b \cr A_8^f \cr B_8 \cr C_8 \cr E_8 \cr F_8 \cr}
\eqn\keightthermmixS
$$
and
$$   \pmatrix{A_8^b \cr A_8^f \cr B_8 \cr C_8 \cr E_8 \cr F_8 \cr}
   \,[T] ~=~ \pmatrix{
      1 & 0 & 0 & 0 & 0 & 0 \cr
      0 & 1 & 0 & 0 & 0 & 0 \cr
      0 & 0 & -1 & 0 & 0 & 0 \cr
      0 & 0 & 0 & -i & 0 & 0 \cr
      0 & 0 & 0 & 0 & e^{3\pi i/4} & 0 \cr
      0 & 0 & 0 & 0 & 0 & e^{-\pi i/4} \cr }\,
   \pmatrix{A_8^b \cr A_8^f \cr B_8 \cr C_8 \cr E_8 \cr F_8 \cr}~.
\eqn\keightthermmixT
$$
These matrices of course contain the $(A_8,B_8,C_8)$ mixing matrices
\ksectfourKeightmixings\ as the appropriate submatrices, and
these six quantities are again modular {\it functions}\/ rather than
modular {\it forms}\/ due to the presence of the tachyonic sixth
quantity $F_8$.
To compensate for this, we introduce the corresponding six-dimensional
$R_\vartheta$ representation with weight $k= -1$
$$ R_\vartheta~\equiv~ \left\lbrace
     {{\thetatwo}\over {\eta^3}},\,
     {{\thetatwo}\over {\eta^3}},\,
     0,\, 0,\,
   \half\left({{\thetathree-\thetafour}\over{\eta^3}}\right),\,
   \half\left({{\thetathree+\thetafour}\over{\eta^3}}\right)
     \right\rbrace
\eqn\kRthetaeight
$$
and construct the tachyon-free representation $R'\equiv
(R_c-R_\vartheta)\otimes
(R_c-R_\vartheta)$.  It can be proven that each member of $R'$
is a modular form of even weight $k= -2$ with respect to a congruence
subgroups at level $N\in 8\bZ$, whereupon the identities
\pbfspliteightids\ immediately follow.

The $K=16$ case is similar.  Here the expressions
$A_{16}^b$ and $A_{16}^f$ are members of the five-dimensional representation
$R_c\equiv\lbrace A_{16}^b,A_{16}^f,C_{16},E_{16},F_{16}\rbrace$
[where these five quantities are defined in \psixteenthermodefs];
under $[S]$ and $[T]$ these quantities mix as follows:
$$  \pmatrix{ A_{16}^b \cr A_{16}^f \cr  C_{16} \cr E_{16}\cr F_{16}\cr}
    \,[S]~=~
    {{e^{i\pi/4}}\over{2\sqrt{2}}} \,\pmatrix{
      1 &  -1 &  2 &  -2 & 2 \cr
      -1 &  1 &  -2 &  -2 & 2 \cr
      1 &  -1 &  -2 &  0 & 0 \cr
      -1 &  -1 &  0 &   \sqrt{2} &  \sqrt{2} \cr
      1 &  1 &  0 &  \sqrt{2} & \sqrt{2} \cr }\,
  \pmatrix{ A_{16}^b \cr A_{16}^f \cr  C_{16} \cr E_{16}\cr F_{16}\cr}~
\eqn\ksixteenthermomixS
$$
and
$$  \pmatrix{ A_{16}^b \cr A_{16}^f \cr  C_{16} \cr E_{16}\cr F_{16}\cr}
    \,[T]~=~ \pmatrix{
      1 & 0 & 0 & 0 & 0 \cr
      0 & 1 & 0 & 0 & 0 \cr
      0 & 0 & -i & 0 & 0 \cr
      0 & 0 & 0 & e^{7\pi i/8} & 0 \cr
      0 & 0 & 0 & 0 & e^{-\pi i/8} \cr}\,
  \pmatrix{ A_{16}^b \cr A_{16}^f \cr  C_{16} \cr E_{16}\cr F_{16}\cr}~.
\eqn\ksixteenthermomixT
$$
These matrices of course contain the $(A_{16},C_{16})$ mixing matrices
\pSsixteenagain\ as the appropriate submatrices.
Once again we find that only the fifth quantity $F_{16}$ is tachyonic, and
again there exists an appropriate compensating
five-dimensional $R_\vartheta$ representation:
$$ \eqalign{
      R_\vartheta~&\equiv~ \left\lbrace
      \sqrt{ {\thetatwo\over{2\eta^3}} }\, , ~
      \sqrt{ {\thetatwo\over{2\eta^3}} }\, , ~
         0 \, , ~
      \half \left(\sqrt{ {\thetathree \over{\eta^3}} }\,-\,
                  \sqrt{ {\thetafour \over{\eta^3}} }\,\right)\, , ~
      \half \left(\sqrt{ {\thetathree \over{\eta^3}} }\,+\,
                  \sqrt{ {\thetafour \over{\eta^3}} }\,\right)\, \right\rbrace
           \cr
   & ~ \cr
  &= ~ \left\lbrace  c^1_1\,,~c^1_1\,,~0\,,~c^2_0\,,~c^0_0\,\right\rbrace
      ~\cr}
\eqn\kRthetasixteen
$$
where in the second line the string functions are at level $K=2$.
This representation $R_\vartheta$ transforms under $[S]$ and $[T]$
with the same mixing matrices as $R_c$;  in particular under $[S]$
the $K=2$ string functions satisfy
$$  \pmatrix{ c^0_0 \cr c^2_0 \cr c^1_1 \cr}\,[S] ~=~
      {{e^{i\pi/4}}\over{2}}\,\pmatrix{
        1 & 1 & \sqrt{2} \cr
        1 & 1 & -\sqrt{2} \cr
        \sqrt{2} & -\sqrt{2} & 0 \cr}\,
    \pmatrix{ c^0_0 \cr c^2_0 \cr c^1_1 \cr}~.
\eqn\kcKtwomixings
$$
Since only $c^0_0$ is tachyonic (and in fact has the same tachyonic
terms as $F_{16}$),
the entire representation $R\equiv R_c-R_\vartheta$
is tachyon-free.  The identities \ksixteenthermoids\
then follow by building the tensor-product representation
$R'\equiv R\otimes R\otimes R \otimes R$ and considering congruence subgroups
with levels $N\in 16\bZ$.

Finally, we note that this second series of identities
can also be proven in a slightly different manner.
Once we have constructed the appropriate
representations $R_c$ and $R_\vartheta$ for each $K$,
rather than prove $R_c-R_\vartheta=0$
it is possible to instead demonstrate $R_c/R_\vartheta =1$
(where $R_c/R_\vartheta$ denotes the representation formed by
dividing each member of $R_c$ by the corresponding member of $R_\vartheta$).
It is clear that $R_c/R_\vartheta$ is itself a valid representation of
the modular group with modular weight $k=0$, and by the usual
arguments it is straightforward to demonstrate that each member of
this ``quotient'' representation is, for example, $\in M_0[\Gam']$ for
an appropriately chosen $\Gam'\subset \Gam$.
 From the theorem presented in the previous subsection it then
follows that each member of this quotient representation
is a constant, and one can verify
from the $q^0$ term in a $q$-expansion of each that this constant is $1$
(because the relative normalization of $R_c$ and $R_\vartheta$ is fixed
by the requirement that they have the same tachyonic structure).
\vfill\eject

 {\bf \underbar{Third series -- the $[K,1]$ identities:}}~~

This series of identities is actually the simplest to prove.
Recall that the Dedekind $\eta$-function satisfies
$$   \eta\,[S]~=~ e^{-i \pi/4}\,\eta~,~~~~
     \eta\,[T]~=~ e^{i \pi/12}\,\eta~,
\eqn\ketatransrecall
$$
and that each identity in this series is of the form
$$    \eta^p \, Q_K~=~ 1~
\eqn\kthirdseriesform
$$
where $p$ is a given power and
$Q_K$ is a sum of terms each containing $p$ factors of level-$K$
string functions.
Since in each case $Q_K$ satisfies
$$   Q_K \,[S]~=~ e^{i p \pi/4}\, Q_K~,~~~~
   Q_K\,[T]~=~ e^{-i p \pi/12}\, Q_K ~,
\eqn\kcsatisfy
$$
and since in each case the product $\eta^p Q_K$ is tachyon-free,
it follows that
$$    \eta^p\,Q_K~\in~M_0[\Gam] ~=~{\bf C}~.
\eqn\kconclusion
$$
Overall normalizations have been chosen in each case such that this
constant is always $1$.
Note that this proof yields the familiar $\vartheta$-function identity
$\thetatwo\thetathree\thetafour=2\eta^3$ in the $K=2$ special
case, and just as easily yields
its higher $K>2$ string-function generalizations.
Thus, we see once again how all of our series of identities
provide the natural generalizations of their known $K=2$ special cases.

\endpage

\ACK
It is our pleasure to thank C.~Callan, C.S.~Lam, A.~Shapere,
C.~Vafa, and E.~Witten for useful discussions and suggestions.
K.R.D.~also thanks the High-Energy Theory Group at Cornell University
for its hospitality while portions of this work were carried out.
This work was supported in part by the U.S.~National Science
Foundation, the Natural Science and Engineering Research
Council of Canada, and les Fonds FCAR du Qu\'ebec.

\endpage
\refout
\bye